\documentclass[12pt,preprint]{aastex}

\def\gs{\mathrel{\raise0.35ex\hbox{$\scriptstyle >$}\kern-0.6em
\lower0.40ex\hbox{{$\scriptstyle \sim$}}}}
\def\ls{\mathrel{\raise0.35ex\hbox{$\scriptstyle <$}\kern-0.6em
\lower0.40ex\hbox{{$\scriptstyle \sim$}}}}
\def\subsun{$_{\odot}$}

\shortauthors{Owen et al}
\shorttitle{Deep Survey of A2125 II.}

\begin{document}

\title{A Deep Radio Survey of Abell 2125 II.}
\title{Accelerated Galaxy Evolution during a Cluster-Cluster Merger}

\author{
Frazer\,N.\ Owen,\altaffilmark{1,2}, 
M.\,J.\ Ledlow\altaffilmark{3,2,4},
W.\,C.\ Keel,\altaffilmark{5,2},
Q.\, D.\ Wang\altaffilmark{6},
G.\,E.\ Morrison,\altaffilmark{7,2}
}
\altaffiltext{1}{National Radio Astronomy Observatory, P.\ O.\ Box O,
Socorro, NM 87801 USA.; fowen@nrao.edu The National Radio Astronomy
Observatory is
facility of the National Science Foundation operated under cooperative
agreement by Associated Universities Inc.}
\altaffiltext{2}{Visiting astronomer, Kitt Peak National Observatory,
National Optical Astronomy Observatories, operated by AURA, Inc.,
under cooperative agreement with the National Science Foundation.}
\altaffiltext{3}{Gemini Observatory, Southern Operations Center, AURA,
Casilla 603, La Serena, Chile}
\altaffiltext{4}{Deceased 5 June 2004. We shall miss his cheerfulness,
unfailing good sense, and scientific industry.}
\altaffiltext{5}{Dept. of Physics \& Astronomy, University of Alabama,
Tuscaloosa, AL 35487 USA}
\altaffiltext{6}{Department of Astronomy, University of Massachusetts,
710 North Pleasant Street, MA 01003}
\altaffiltext{7}{National Optical Astronomy Observatories, 950 N.
Cherry Ave., Tucson, AZ 85719 USA}

\setcounter{footnote}{7}

\begin{abstract}

	Using  our extensive radio, optical,
near-IR and X-ray imaging and spectroscopy, we consider
the reason for the unusually large number of radio
detected galaxies, mostly found outside the cluster core,
in Abell 2125($z=0.2465$, richness class 4). With 20-cm VLA data,
we detect continuum emission from 90 cluster members. 
The multiwavelength properties of these galaxies suggest
that most of the radio emission is due to an enhanced 
star-formation rate. The dynamical study
of \citet{m03} suggests that Abell 2125 is undergoing a
major cluster-cluster merger, with our view
within 30 degrees of the merger axis
and within 0.2 Gyr of core passage. The combination of
projection effects and the physical processes at work during this
special time in the cluster's evolution seem likely to
be responsible for the unusual level of activity we see in the
cluster. We argue that tidal effects on individual cluster
members, often far from the cluster core, are responsible
for the increased star formation. Our results are
consistent with the idea that disk galaxies during this phase of a 
cluster's evolution undergo rapid evolution, through a burst of
star formation, on their way to becoming S0's.

\end{abstract}


\section{Introduction}

	With the recent convergence of supernova and CMB research to
a remarkable consensus on the geometrical
parameters of the universe, understanding the evolution of galaxies
with cosmological epoch is, perhaps, the major goal of extragalactic 
astronomical research. From the morphology-density relation and the 
Butcher-Oemler
effect, it is clear that the large-scale process of cluster formation
must be a key part of the picture. But understanding galaxy evolution
may be more difficult than geometric cosmology because of the complex
role of dissipational processes, and the confusing results of observing
two-dimensional projections of three dimensional physical
objects (e.g. galaxies and clusters alike). Using
as large a range of wavelengths as possible is one way to improve our
understanding of what we are seeing. In this paper, we describe a
study, using radio, optical, near-IR and X-ray
imaging and spectroscopy, to try to understand how a very rich, very 
radio-active cluster, Abell 2125,  fits into the bigger picture.

	We began our study of the Abell 2125 field with moderately deep 20cm 
VLA C-array observations \citep{dow,o99}.
We compared Abell 2125 (a richness class 4, blue cluster at $z \sim
0.25$) with Abell 2645 (an apparently similar cluster at the same
redshift but with much redder galaxies). We found a much higher detection
rate of radio galaxies (27 vs 4) in Abell 2125, while the detection rate in
A2645 is consistent with a normal AGN population in a lower redshift
cluster. The excess in Abell 2125 occurs entirely in objects with
$L_{20cm} < 10^{23}$ W Hz$^{-1}$, and thus seems consistent
with a star-forming population \citep[e.g.][]{ccb}. However, only
a small fraction of these radio-selected galaxies shows optical
evidence of so much star formation, leaving the interpretation of the
radio results in some doubt.
Either the star formation in most of the low-luminosity radio sources
is very well hidden at optical wavelengths, or we must seek some other explanation 
for the radio activity.

	Closer examination of the optical and X-ray properties of Abell 2125 shows 
that it is not a typical, very rich Abell cluster.
The galaxy distribution shows a central
concentration together with an extension to the southwest, spanning
at least 2 Mpc. X-ray observations \citep{wang}
show the same pattern, and that the total X-ray luminosity is low
for such a rich cluster. \citet{wang} suggested that we might be seeing
diffuse X-ray emission from a superstructure, as is seen in many 
cosmological simulations
of large scale structure formation. Much of the excess radio-emitting
galaxy population from \citet{o99}, like the diffuse X-ray emission, is
contained in the southwest extension, not the cluster core. These factors
suggest that the unusual galaxy population, especially as seen at radio frequencies,
might be connected to the large-scale nature of Abell 2125 itself.

	To study this cluster further, we have made much deeper
and higher resolution radio observations using all four VLA configurations.
These deep radio data have also motivated deep optical and NIR imaging,
optical spectroscopy, deep submillimeter observations, and a long
{\it Chandra} exposure. While some of these ancillary observations were 
motivated by possibility of studying a background sample, we consider here what
they tell us about Abell 2125.

	Paper I \citep{o03a} described our radio,
optical and NIR observational program to study the region of the
sky centered on Abell 2125. In this paper, we now discuss the 
properties of the cluster
members, particularly in light of the unusually large
number of radio detections from cluster's members. 
Physical quantities are calculated using a WMAP $\Lambda$CDM 
cosmology with
$H_0$=71 km s$^{-1}$ Mpc$^{-1}$, $\Omega_M=0.27$ and $\Omega_V=0.73$.

\section{Observations}

\subsection{Ground-based and X-ray Imaging and Spectroscopy}

	Most of the observations and reductions have been
described elsewhere.
	The radio, optical and near IR imaging and reduction is 
described in paper I \citep{o03a}. The optical spectroscopy is 
discussed in \citep{m03}. The {\it Chandra} imaging, reductions
and basic results are presented in \citep{w03}.

\subsection{HST Imaging}

We observed three fields in Abell 2125 with HST and WFPC2, centered
to maximize the number of radio-detected members within the fields
without imposing tight orientation constraints. One field is just
west of the cluster core, with the other two spanning the clump of bright
galaxies 5' SW of the center. Each of these latter two fields was observed
twice; for the southwest field, at significantly different orientation
angles, and for the southern field, at pointings differing by about 10
arcseconds. Both F606W and F814W filters were used, corresponding roughly
to the emitted $B$ and $R$ bands at $z=0.2465$. Each filter was used
for an exposure of 2600 seconds per observation, split into two equal parts
for cosmic-ray rejection.

Several of the images were taken at small angles to the illuminated limb of the 
Earth, and suffered from scattered Earthlight. This produces a diagonal
X pattern of brighter regions. These were removed following the
same basic precepts as used for the original Hubble Deep Field data
\citep{wil96}. Where multiple pointings with similar 
scattering properties were available, they were combined to reject actual 
objects, then scaled to a common scattering intensity and smoothed. This 
scattering model was applied, with appropriate scaling, to all the affected 
images. Since this procedure does not necessarily preserve mean sky brightness
as it affects photometric errors, nor impose a consistent sky level across
the various CCD fields, we added a constant value to give the same level
for all chips.

\section{Results}

	In Table~\ref{SL}, we present our basic observational
results for spectroscopically confirmed, radio detected members of
 Abell 2125. In column 1 we list radio source name; columns 2 and
3 contain the optical R.A. and Declination (2000.0)
corresponding to the galaxy (for members identified with radio emission detected
at $ > 5 \sigma$ above the noise). Column
4 contains the distance in arcmin from the nominal cluster center
[15 41 15, 66 16 00] \citep{o99}. In columns 5 and 6, we list the peak 
and total
flux density for each radio source in $\mu$Jy. Column 7 contains the
size (or a limit) in arcsec. If the source is resolved and fitted with
a two dimensional Gaussian, we give the deconvolved major axis size
(FWHM), 
the minor axis size and the position angle. If only a size is given,
this value has been estimated from the image by hand. Column 8 lists
our best measured value for the redshift; an ``e'' indicates
that emission lines were present in the spectrum \citep{m03}.

\subsection{Photometric Analysis}

	We have quantified the spectral energy distribution ,SED,
of each radio galaxy, using the publicly-available program {\it hyperz} \citep{bol00} to
fit template spectra. We measured optical magnitudes for each of
our 10 bands from U to K in the Gunn-Oke aperture (radius 13.1 kpc).
In table~\ref{data}, we give AB magnitudes used for the {\it hyperz}
results quoted. -9.00 in table~\ref{data} indicates no measurement
was available. 
Then {\it hyperz} was used, with the redshift range constrained to fall
within the cluster range in light of the spectroscopic redshifts, to fit the 
Bruzual-Charlot models provided
with the program  \citep[GISSEL98;][]{bc93} with
 a Calzetti extinction law \citep{c00} and with a total age 
less than the age of the universe at $z=0.2465$, about $10.8$ Gyr. 
The result yields a spectroscopic 
``galaxy type'', an age for the dominant star-forming event, an
$A_V$ extinction for the Calzetti law, and a rest-frame, R absolute
magnitude. In table~\ref{T}, we summarize the model star-formation histories used. 
The galaxy type is based on an assumed $e$-folding timescale
for the star formation. This
timescale, combined with the age assigned to each template, determines
each SED in the catalog. For ages which are a
significant fraction of a Hubble time, templates for different
galaxy types may not differ very much. It is not clear that
the galaxy type assigned to each $e$-folding timescale is closely
correlated with the galaxy morphology. Finally,
the galaxy ages are a (luminosity-weighted) global estimate for the stars in the 
aperture and may not reflect the age of the star-formation
event which we are observing in the radio. Thus we prefer
to look
on the types as a way of parameterizing the catalog with a useful
range of assumed star-formation histories. The dominant parameter in a template
is the age of the assumed
stellar population, especially
for ages of a Gyr or less. The extinction is a useful dimension
to explore and seems generally correlated with star-formation,
although derived age and extinction are somewhat anti-correlated. The absolute
magnitude should be robust, since we are essentially using
{\it hyperz} to interpolate between our measured magnitudes using the
best fitting SED. 

\subsection{Emission Lines}

    Another important clue to the nature of radio population
is the detection (or lack) of emission lines. Our spectroscopy
comes from several different observing runs with different wavelength
ranges covered. Thus we cannot report a common set of emission
lines. However, in almost all cases, we detect [OII] and/or
H$\alpha$/[NII]. Often [OIII], H$\beta$ and [SII] are detected.
As reported
in \citet{m03}, almost all the galaxies with detected emission lines
have line ratios indicating that they are powered by star formation. Of
course, for very powerful AGNs, one expects to see a rich spectrum
of broad and/or narrow lines. However, at the radio luminosities and
redshift we are dealing with, one rarely can detect emission lines
in contrast against the stellar continuum \citep{olk}. Thus for
the relatively low-luminosity radio objects we are observing, the
lack of detected emission lines together with detected
radio emission is consistent with (weak) AGN activity.
However, the lack of emission lines can also be consistent with
significant star formation if dust obscures the line
emission \citep{m02}. Thus when we see line emission in these systems, that
result normally indicates star-formation activity as the origin
of the radio emission. However, for the galaxies in which we do not detect line emission,
the situation is less clear.

\subsection{Physical Parameters}

	In table~\ref{pp} we summarize the physical parameters for
each galaxy with radio emission. All values have been calculated
assuming a redshift of 0.2465 \citep{m03}.
Column 1 contains the source
name; column 2, the largest linear size; column 3, the
absolute radio luminosity; column 4, the projected distance
from the cluster center; column 5, the absolute, rest frame
R magnitude, column 6, 7 and 8, the {\it hyperz} results: spectral
type (table~\ref{T}), age and $A_V$, and column 9, the spectroscopic redshift.
Once again an ''e'' appended to the redshift indicates emission lines
are detected. 

\subsection{{\it Chandra} Detections of Cluster Members}

    Compact X-ray sources in galaxies are usually
associated with active galactic nuclei (AGN). Our {\it Chandra} data for Abell 2125 reach
flux levels at which star-forming galaxies can also be detected 
\citep{ran03}. On the other hand, detection of X-ray emission on
larger scales than galaxies is usually associated with free-free
emission from the hot cluster medium. Thus one major goal of the 
{\it Chandra} observation was to use X-ray detections of compact
emission from the individual galaxies with radio emission
to recognize AGN. If a large fraction of the radio activity was
driven by AGN, even in emission-line galaxies, we might expect
to detect a large fraction of the radio-emitting cluster members.
The {\it Chandra} observation shows 99 discrete sources, as
detailed in \citet{w03}.

	Eight radio-detected galaxies in Abell 2125 also host point-like
X-ray sources independently detected from the {\it Chandra} observation (table~\ref{XR}). One of these
sources, 00047 (also called C153) shows unique extended structure on close
inspection, as discussed in detail in our paper on the cluster
core \citep{o03b}. However, the low background level of the {\it Chandra}
detectors allows deeper detections of sources if the positions are independently and
accurately known; using the optical positions to fix locations of potential
X-ray sources yields an additional twelve X-ray detections
with probabilities $< 10^{-3}$ of being random events.  The true
probability of these weaker detections being spurious was further evaluated
by shuffling the coordinates of the galaxies and repeating the search.
These tests indicated that $\sim 0.8$ detections could be expected
by chance for the entire radio galaxy sample from this procedure.
The errors on the fluxes are large for these detections and there
is a statistical bias to overestimate the fluxes by as much as a
factor of two \citep{wang04}. Thus the luminosities quoted should
be taken only as very rough estimates. These additional 12 faint
X-ray detections are listed in  table~\ref{FXR}. Once again one of the radio detections 
(00057) is clearly extended in X-ray emission, associated with one of members of the 
central triple in the
cluster core (although it is not detected by the point source
detection algorithm used by \cite{wang04}). This galaxy is also
discussed in \citet{o03b}. 

	In table~\ref{XNR} we list two other cluster members which
are associated X-ray sources reported by \citet{w03}. Neither of
these galaxies has detected emission lines. However, one (X072)
also has a marginal radio detection, with a peak between 3--4 $\sigma$ 
on the radio image. The other source (X064), associated with a cD-like
galaxy, is also extended and soft.

	The new X-ray analysis brings the detected fraction among
radio-selected cluster members to 20/73, within the region covered by the {\it Chandra} image.
Excluding the very extended sources in the cluster core, all
the detections by \citet{w03} are associated with relatively
bright galaxies ($M_R \le -22.0$) without emission lines. Three
of the weaker X-ray objects from table~\ref{FXR} also lack emission lines
and only one is fainter than $M_R = -22.4$ (i.e. $-21.8$). All of
these objects seem consistent with AGN. Furthermore, the 
X-ray morphology of five of the additional, weaker objects associated
with emission lines (and one without emission lines) are not clearly 
point-like (at 6 arcsec or 23 kpc resolution); therefore these X-ray
sources are large with respect to the galaxy scale and  these galaxies may
simply be confused with a local peak in the extended X-ray 
emission. Thus it is hard to
evaluate the origin of the the X-ray detections for these objects.
Only three of the weaker galaxies with emission lines do appear
point-like. These objects (00022, 00026 and 00027) are associated with 
relatively radio-luminous objects consistent with star formation.
The X-ray luminosities of these objects also are consistent with 
star formation at the same general level as implied by the radio
emission within the errors \citep{ran03}. 

      In summary, 53/73 radio-detected  galaxies in the {\it Chandra}
field are {\it not} detected in X-rays. Ten of the X-ray-detected objects are
consistent in optical, radio, and X-ray properties with AGN. Three others are 
most consistent with star formation strong enough for X-ray detection. The rest are either 
associated with diffuse
emission, or are too weak and/or in regions too complex to characterize
clearly. Thus the {\it Chandra} data do not suggest that the bulk of
the radio activity is due to AGN.  

\subsection{Radio Luminosity/Morphology}

    Statistically, $10^{23}$ W Hz$^{-1}$ is the approximate crossing
point of the luminosity functions of radio galaxies driven by AGN
and star formation. Above $10^{23}$ W Hz$^{-1}$ AGN
dominate, below $10^{23}$ W Hz$^{-1}$ star-forming galaxies dominate
\citep{ccb}. In rich clusters, where there are more early-type
galaxies, the log of the crossover luminosity is nearer 22.8
\citep{m02}. However there are, of course, objects above and
below the crossover luminosity driven by the non-prevailing mechanism.
Thus radio luminosity is a {\it statistical} indicator of AGN or
star-formation-driven radio emission. 

   Radio morphology is another important indicator. Clear, FR I or
FR II morphology is a clear AGN indicator. Most of the sources
with radio luminosities above $10^{23}$ W Hz$^{-1}$ in A2125 show
clear FR I or FR II morphology \citep{dow}.

\subsection{Radio/Optical Alignment}

    Beyond source luminosity, another important indicator that radio emission is driven
by star formation is alignment of radio emission with the
major axis of a galaxy. While star formation in some
systems is too concentrated in the galaxy core to resolve at
our 1.5" (5.7 kpc) limit, extended and aligned radio
emission is a strong clue that the emission is driven by
star formation.

\begin{figure}
\plotone{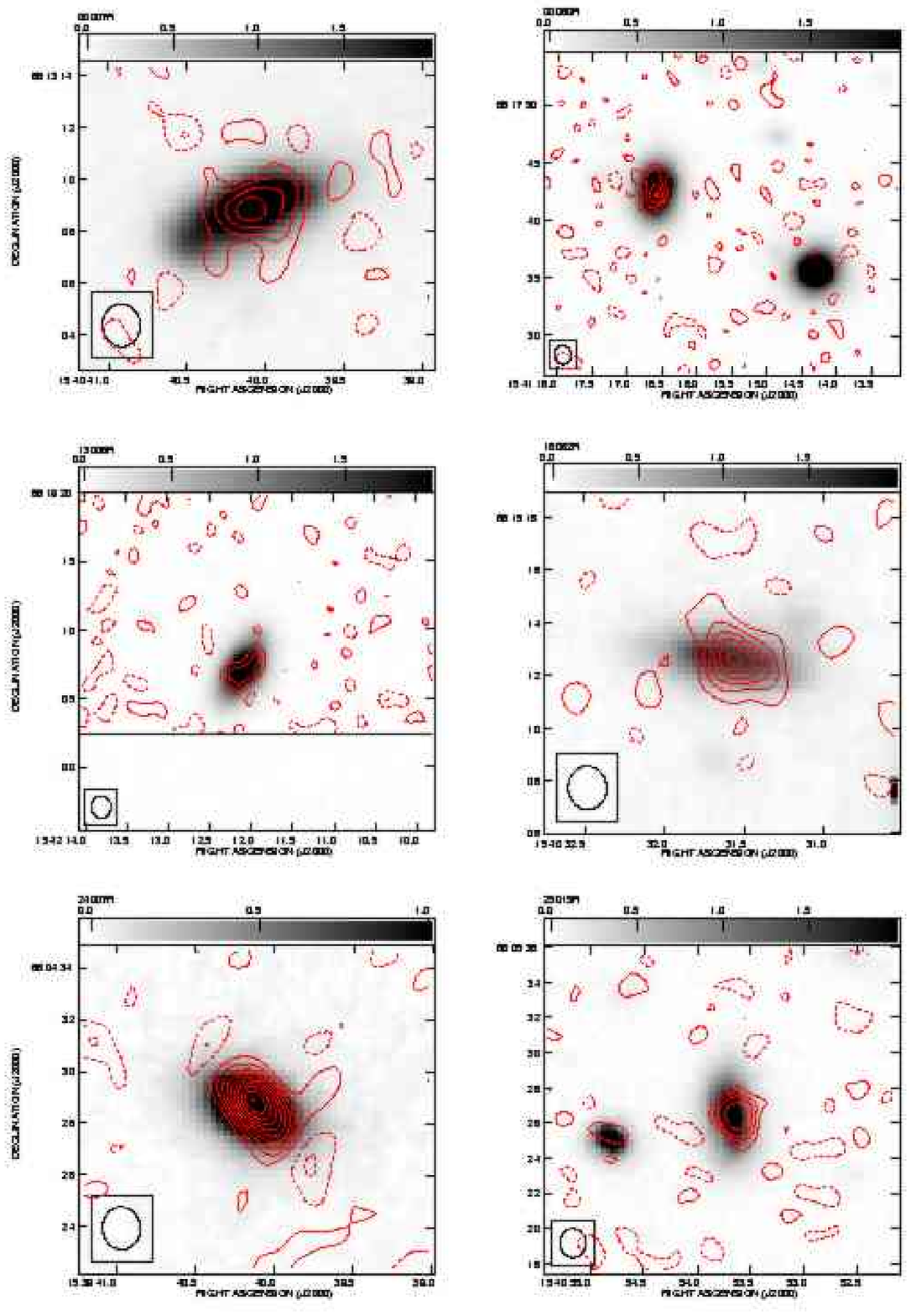}
\caption{Examples of Radio Galaxies with Aligned Radio/Optical
Emission. \label{RO1}}
\end{figure}

\begin{figure}[p]
\plotone{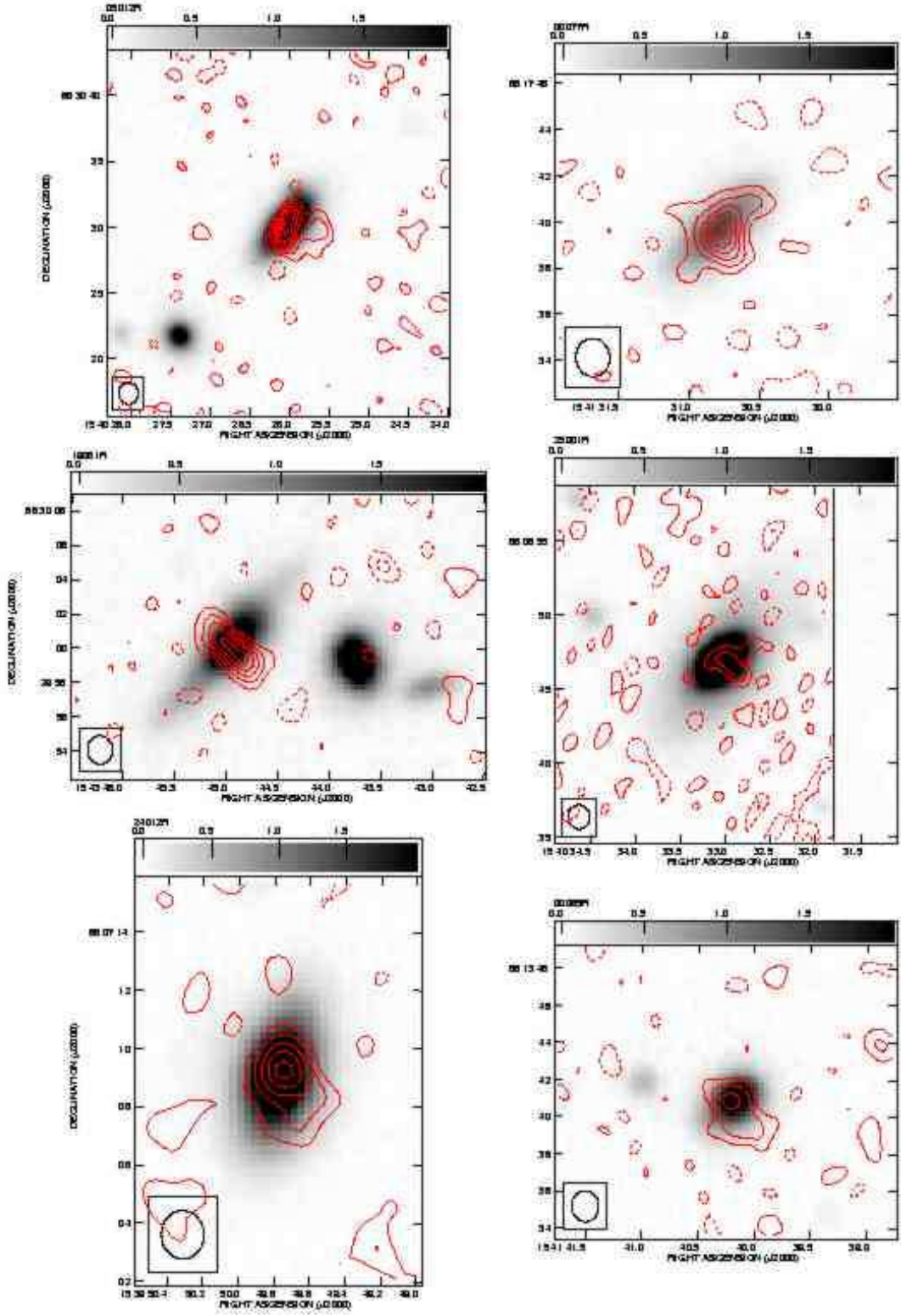}
\caption{Examples of Radio Galaxies with Some Misaligned Radio/Optical
Emission. \label{RO2}}
\end{figure}

	 A large
fraction of the lower-luminosity sources show alignment between
the radio and optical structures. In figure~\ref{RO1} we show
radio contour maps overlaid on the MOSAIC R-band optical images
as examples of the aligned structures. However, some of the radio
cluster members show other features which are not aligned. In
many cases, both aligned and misaligned radio structure can be
seen. In some cases only misaligned structures
are seen. In figure~\ref{RO2} we show some of these
misaligned features. Some of the misaligned radio features could
represent outflows directed along the minor axis, either the jets of AGN, or 
starburst-driven winds
\citep{co96}. In some cases, these features could be evidence
of on-going stripping of the galaxy's ISM as it moves relative to
the external medium. 

\subsection{Concentration Indices}

Optical morphology is another important clue to understanding
what we are seeing in galaxies. However, except in the small regions imaged with HST, 
the resolution of our optical imaging,
is too low for detailed classification.
We therefore quantify the morphological types spanned by the radio
galaxy sample using concentration indices measured for 81/90 of the
sources.  No reliable measures were obtained for the other 9 galaxies
due to the proximity of brighter companions or very bright stars.  

We measure the concentration index (similar to a bulge-to-disk ratio),
defined as the ratio of the flux between inner and outer isophotes,
where the outer isophote is measured to 24.75 magnitudes arcsec$^{-2}$
in the rest-frame of the galaxy. The inner isophote is normalized to
0.3 times the radius of the outer isophote. While our imaging is
sufficiently deep to extend to fainter limiting surface-brightness
(\citet{a94} used 25.5 mag arcsec$^{-2}$), the majority of
the radio galaxies in A2125 are in high galaxy-density regions, many with close
companions, making it difficult to separate overlapping isophotes.  The 24.75
mag arcsec$^{-2}$ limit was a compromise which allowed us, with masking, to fit
nearly all the galaxies to a uniform depth. The difference in CI measure with
different surface-brightness limits is fairly small (of order
0.02-0.03), which is an estimate of the typical error bars. 

\begin{figure}
\plotone{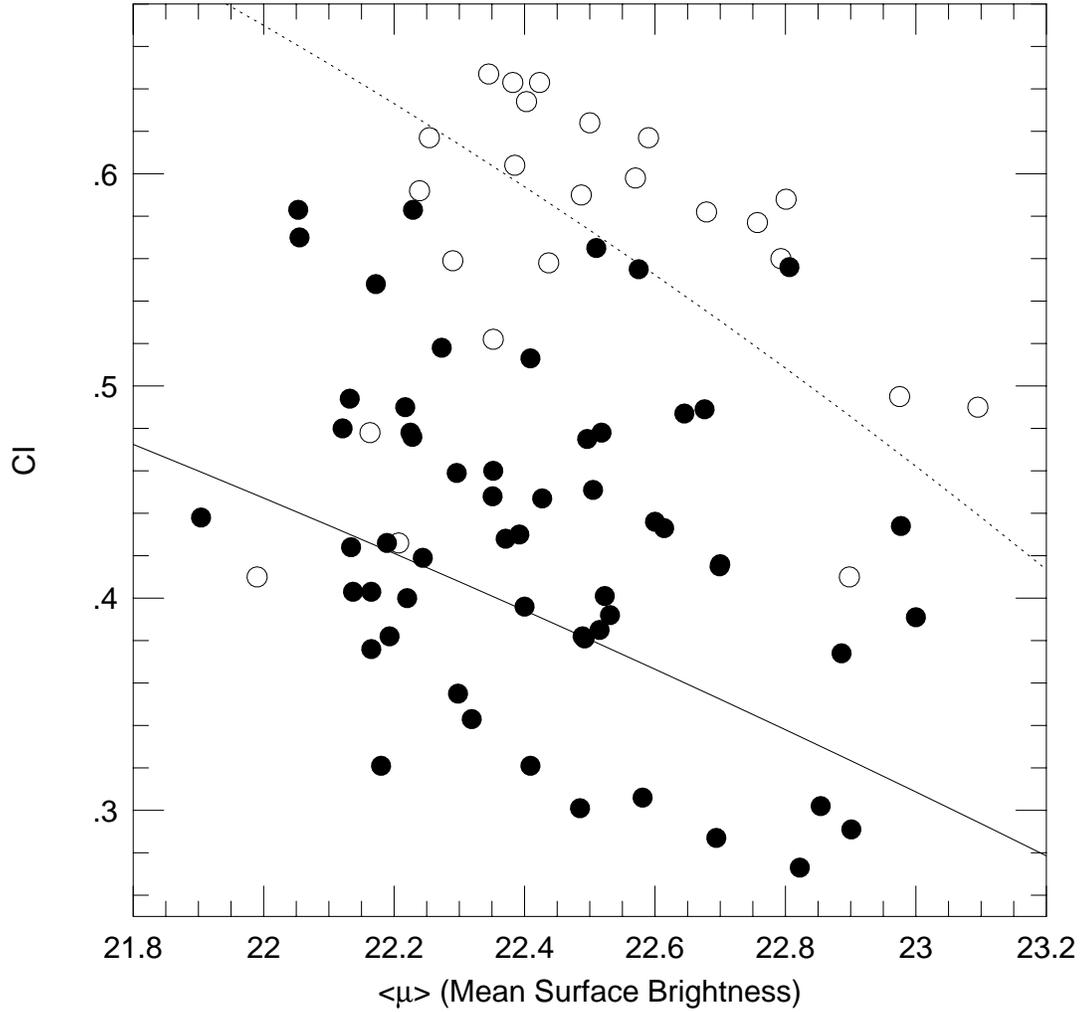}
\caption{Concentration Index for the radio detected galaxies in
A2125 plotted versus Mean Surface Brightness. The theoretical
curves for $r^{1/4}$ laws (dotted line) and exponential disks
(solid line) are also plotted. The solid symbols are galaxies
which has a Star Formation Index, $SFI > 1$, indicating some
evidence of active star-formation. The open circles have
$SFI \le 1$ indicating little or no evidence for star-formation
activity.
 \label{CI}}
\end{figure}

In figure~\ref{CI} we plot the concentration index versus mean 
surface-brightness for 81 cluster radio galaxies. The upper
dotted line shows the expected location of galaxies with $r^{1/4}$
laws, while exponential disks are plotted with a solid line. There is clearly a
spread in values, with
a substantial (30/81, 37\%) fraction falling intermediate between
bulge and disk-dominated profiles, as would be characteristic of an S0
classification. The open circles in figure~\ref{CI} indicate objects in which 
the SFI, discussed in the next section, suggests little or
no evidence of star-formation based on a number of indicators.  The solid
symbols indicate some evidence of star formation based on SFI. Clearly
most of the objects near the $r^{1/4}$ law line are consistent with
old stellar populations while most of the points below this line show
some evidence for star formation.

\subsection{Star formation or AGN ?}

	One of the major questions raised by the galaxy
population in Abell 2125 is whether the excess radio
emission is due to star-formation or nuclear activity.
As discussed in  section 3.4, the X-ray results, perhaps
the most robust single tracer of AGN.
show that less than 15\% of the radio sample appear to 
be AGN.
As discussed in \citet{o99}, for cluster of this richness, 
there is no excess in the radio-galaxy
population above $10^{23}$ W Hz$^{-1}$ where
radio AGN normally dominate. However, many of the optical
identifications below  $10^{23}$ W Hz$^{-1}$ do not
show the obvious colors and emission lines one might
expect from star-bursting galaxies. For galaxies
at the distance of Abell 2125, mainly observed from the ground,
any given star-formation indicator may or may not be
detectable depending on the details of an individual
galaxy's properties. With the relatively coarse spatial sampling
used for optical imaging and spectroscopy, much of the
light from a massive galaxy may come from its old stellar
population. The orientation and dust content will 
also affect the emission line equivalent widths and the effective
age of the SED. For systems close to edge-on, the
radio emission from star formation is expected to
align with the galaxy disk, if it is resolved. But
for a face-on system, this alignment is harder to
detect. In practice, we need to look at several
different indicators to understand the likely origin
of the radio emission.  

	With our new dataset, much larger 
than used in \citet{o99},
we can look for evidence of star formation in several
different ways. Indicators we can use include: 1) radio luminosity,
2) radio alignment with optical galaxy major axis, 3) optical
emission-lines, 4) optical/NIR SED, 5) dust $A_V$ required
by the best fit SED, 6) absolute R-magnitude. We calculate
a weighted ``star-formation index'' (SFI) for each galaxy to give us 
an overall indication
of the likelihood of star formation being responsible for
the radio emission. Some indicators seem better than others, so
we will assign weights of 0.5 or 1.0 for each of the six properties depending
on how good the evidence seems to be. Radio alignment and
emission lines seem to be particularly good indicators, so each of
these is worth 1.0 in our scheme. Radio luminosity is only a
statistical indicator and in clusters, is really only a strong one when
$\log(L_{20cm}) < 22.7$, so we add 0.5 in this case. Likewise,
\citet{lo96} and \citet{M01} show that AGN
in clusters are rarely if ever more than one magnitude fainter than
$L_{*}$. This limit corresponds to $M_R=-21.0$ for our magnitude
system. We will add 0.5 to the score for  $M_R$ fainter than $-21.0$. 
Likewise most AGN are associated with old stellar populations and even
strongly star-forming galaxies can appear to have relatively
old SEDs. Thus we will add 0.5 to the score if the age of the
best fit SED is $< 5$ Gyrs and 1.0 if it is $< 1$ Gyr. Finally,
dust correlates with star formation. We will add 0.5 for an
$A_V > 0.5$ and 1.0 for  $A_V > 1.0$. Thus the maximum score
in this system is 5.0. 

	Table~\ref{SFI} summarizes the SFI scores for the 90
galaxies with radio detections. 
Only one galaxy has a perfect 5.0 rating, while 21/90 have SFI $ < 1$.
These seem likely to be AGN. Fifteen have intermediate SFI of 1 or
1.5. These systems are consistent with star-formation but are not
as certain as galaxies with higher SFI. The 54 with SFI of 2 or more 
and seem likely to have
radio emission driven by star formation. This comparison of
multiwavelength indicators suggests that most of the
excess activity in Abell 2125 is driven by star formation.

\begin{figure}
\plotone{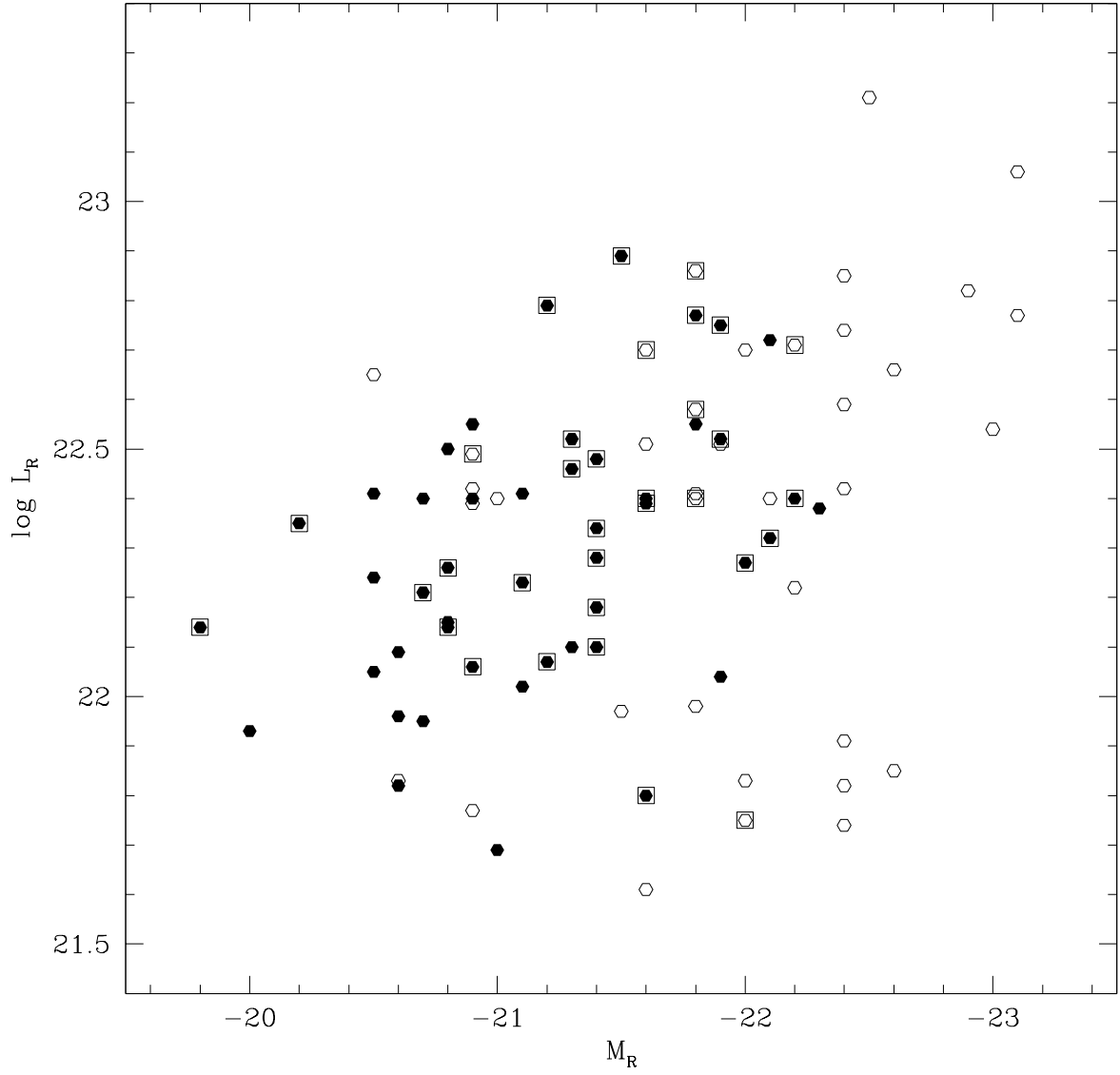}
\caption{R absolute magnitude versus log of the 20cm absolute
radio luminosity (W/Hz). Solid symbols indicate sources with
line emission. Boxes indicate objects for which the radio emission
is aligned with the optical major axis. \label{fig1}}
\end{figure}

	In figure~\ref{fig1}, we plot
of the radio luminosity versus the absolute R magnitude. The
solid symbols show which galaxies had detectable line emission;
the squares show which galaxies had aligned radio and optical
emission. We believe these two indicators are the best for 
signifying star formation. Figure~\ref{fig1} suggests
a break point in the galaxy properties at
$M_R = -22.3$. No galaxies brighter than $M_R = -22.3$ show
evidence of star formation, while most of the galaxies fainter
than this show emission lines and/or radio alignment.
Several of the remaining objects fainter than $M_R = -22.3$ also
show dusty SEDs suggesting that we may be losing the emission-line
spectra to dust extinction.

\begin{figure}
\plottwo{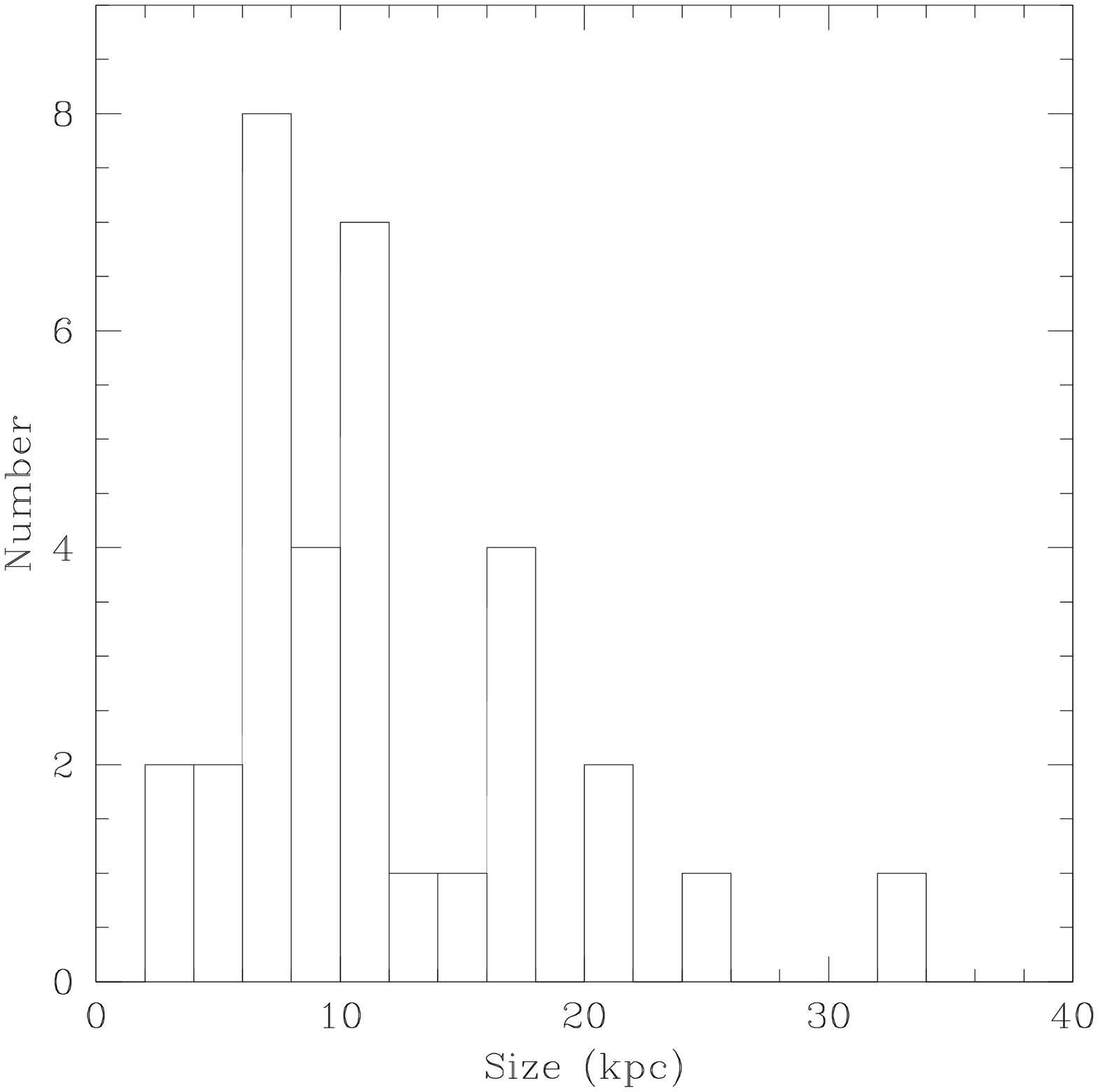}{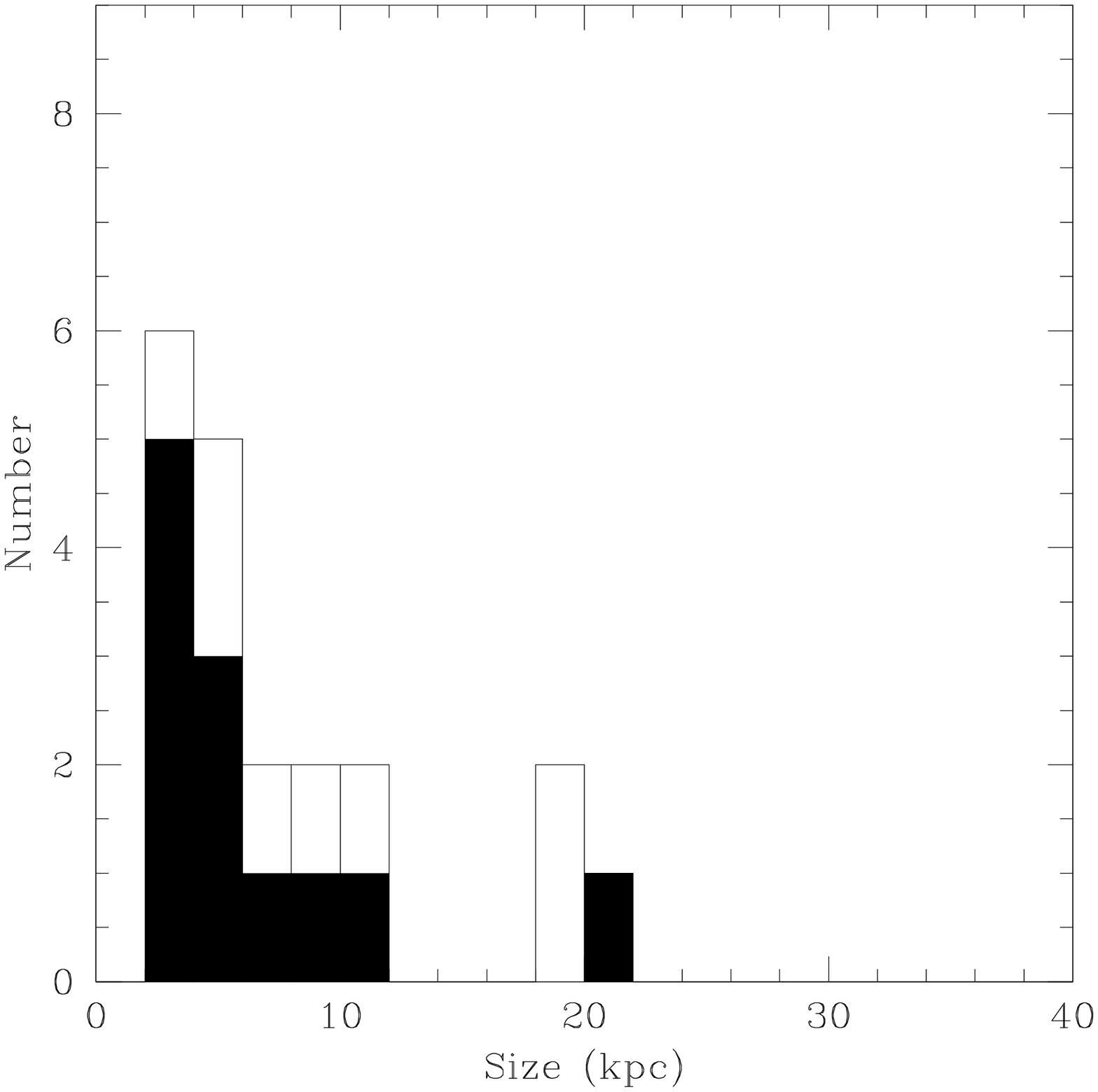}
\caption{Radio linear size histograms. The left panel contains
all radio/optical aligned sources. The right panel includes
only non-aligned sources with emission lines. The solid boxes
represent sources with upper limits to their radio sizes. \label{fig2}}
\end{figure}

 In figure~\ref{fig2} we show histograms of the radio
linear size distributions for the aligned sources (left panel) and the sources
with radio emission without clear alignment (right panel, in which
upper limits are shown in black). Most of the emission-line objects
which show alignment (left panel) have rather extended radio
emission, much larger than the galaxy core. Thus we seem to
be seeing extensive star formation throughout disks of
most of these systems.  Most of the emission-line galaxies
without clear alignment (right panel) have a much smaller radio
extent. Over half only have upper limits to the radio size. These
latter objects
appear to be consistent with more compact star-formation events.
All these objects seem to be consistent with star formation
as the origin of the radio emission.

	 Thus there is a strong tendency for the
optically and radio fainter galaxies to be consistent with
star formation as the origin of the radio emission. Also, by number,
most of the radio-emitting galaxies we have detected are most
consistent with star formation, but almost all galaxies more than 30\% 
brighter than $L_{*}$ are consistent with AGN
activity. 

\begin{figure}
\plotone{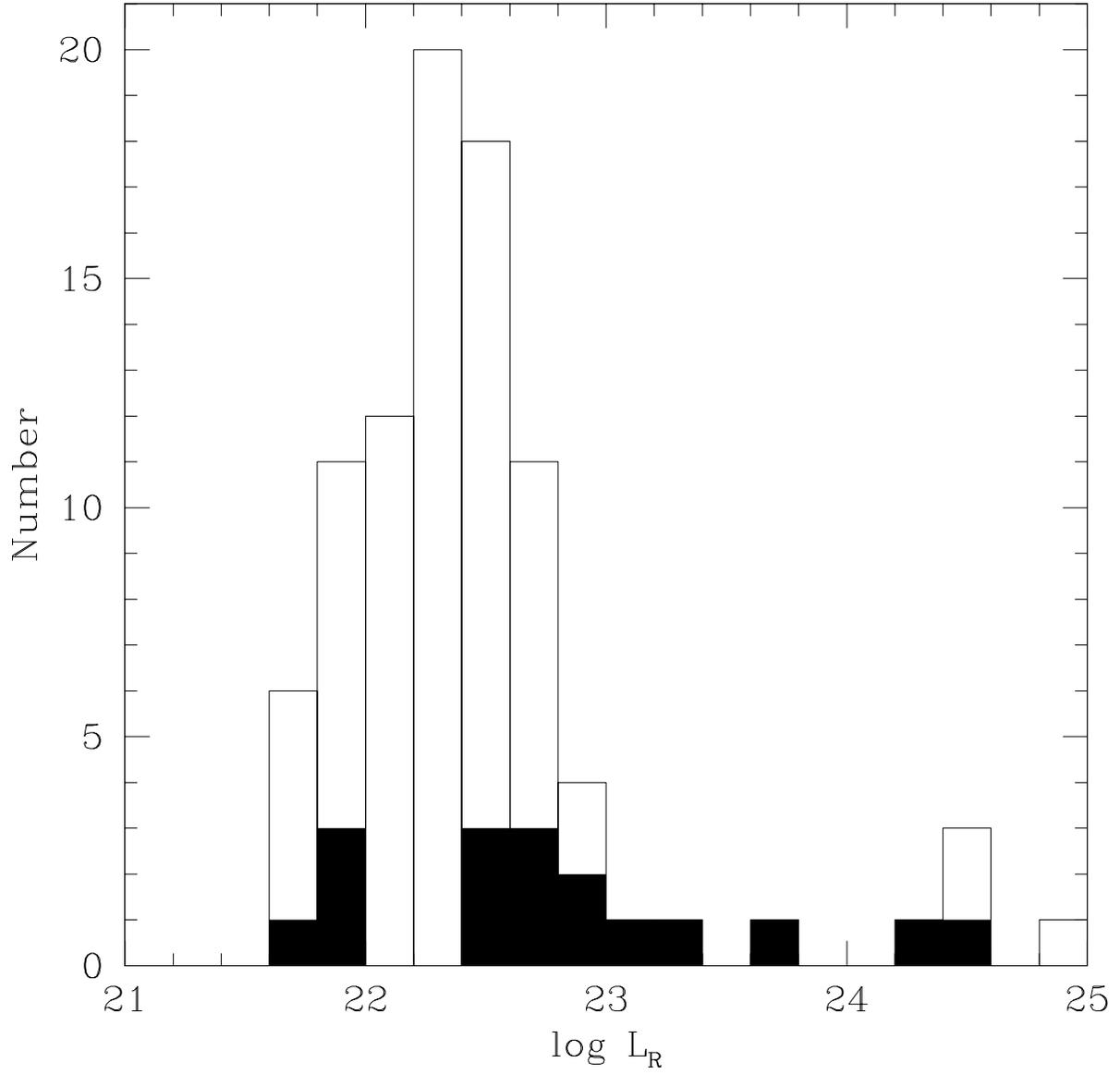}
\caption{Histogram of log of the 20cm absolute radio luminosity.
 Solid boxes indicate objects for which the optical
absolute magnitude is brighter than $ -22.3$. \label{fig3}}
\end{figure}

	In figure~\ref{fig3} we show the histogram for all the absolute radio
luminosities. The black boxes indicate objects with $M_R$  brighter than
$-22.3$. In figure~\ref{fig3} one can see the lower cutoff in 20cm
radio luminosity,
$\log L_R $, at 21.6 and the peak in detections near 22.3. Also one
can see the tendency for the optically brighter galaxies to 
dominate the sample above  $\log L_R > 22.8$, where
one would expect AGNs to dominate as discussed earlier. The drop-off in numbers
at faint levels
probably combines several effects. First, away from the field
center the sensitivity drops off. Second, extended sources at
near the point source limit will be missed. Third, a few faint
objects with fainter optical IDs still do not have redshifts and
could be cluster members. Thus it seems premature to conclude the
the true radio detection rate in the cluster peaks at $\log L_R  > 22$.
A deeper radio survey would likely detect many more star-forming
systems and there is a large population with emission lines in
\citet{m03} which have not yet been detected in the radio. 

	To estimate a SFR implied by the radio luminosities, we
can use the calibration of \citet{yun01}, who used the radio-FIR
correlation to estimate the radio-SFR relation.  \citet{yun01}
estimate the SFR assuming a Salpeter IMF down to
0.1 M\subsun. In order to
include only star formation above 5 M\subsun, we scale their
relation by 0.18, and, corrected to our assumed cosmology, 
the result is
 
\begin{equation}
 SFR (M_{\odot} yr^{-1}) = 1.0\times 10^{-22} L_R (W Hz^{-1}). 
\end{equation}
   
	Thus our peak near $\log L_R  = 22.3$, corresponds to
about a SFR of about 2 M\subsun yr$^{-1}$ (or 11 M\subsun yr$^{-1}$
for a Salpeter IMF down to 0.1 M\subsun). Mostly these galaxies
appear to exhibit relatively modest rates of star-formation. 

	As discussed earlier, while most of the X-ray detections
are most consistent with AGN or diffuse X-ray emission in the
local area surrounding the detections, three of the radio galaxies
have weak point-like X-ray emission. \citet{ran03} have shown that
the X-ray luminosity in star-forming galaxies also correlates
with the SFR. Converting their results to our cosmology and to
the 0.5-8 keV band, assuming a photon number index of 2 \citep{w03}, 
we get

\begin{equation}
 SFR (M_{\odot} yr^{-1}) = 6.8 \times 10^{-41} L_{0.5-8keV}. 
\end{equation}

From equation 1 and the data on the only three radio
detected galaxies in table~\ref{FXR},
one gets radio SFR estimates of 3.2, 3.5 and 5.2 $M_{\odot} yr^{-1}$
for 00022, 00026 and 00027, respectively. From equation 2 the estimates from
the detected X-rays for the same three objects are 6.3, 14 and 14
$M_{\odot} yr^{-1}$, respectively. Given that the X-ray measurements are biased
high, that we are just marginally able to detect
the X-rays from the galaxies with the highest SFR's and that there is 
significant scatter in the radio/X-ray correlation \citep{ran03}, the
agreement in the SFR estimates from both bands is consistent with a 
star-formation origin to both the radio and X-ray emission for these 
three objects.

\subsection{HST Images of Radio Galaxies}

\begin{figure}
\plotone{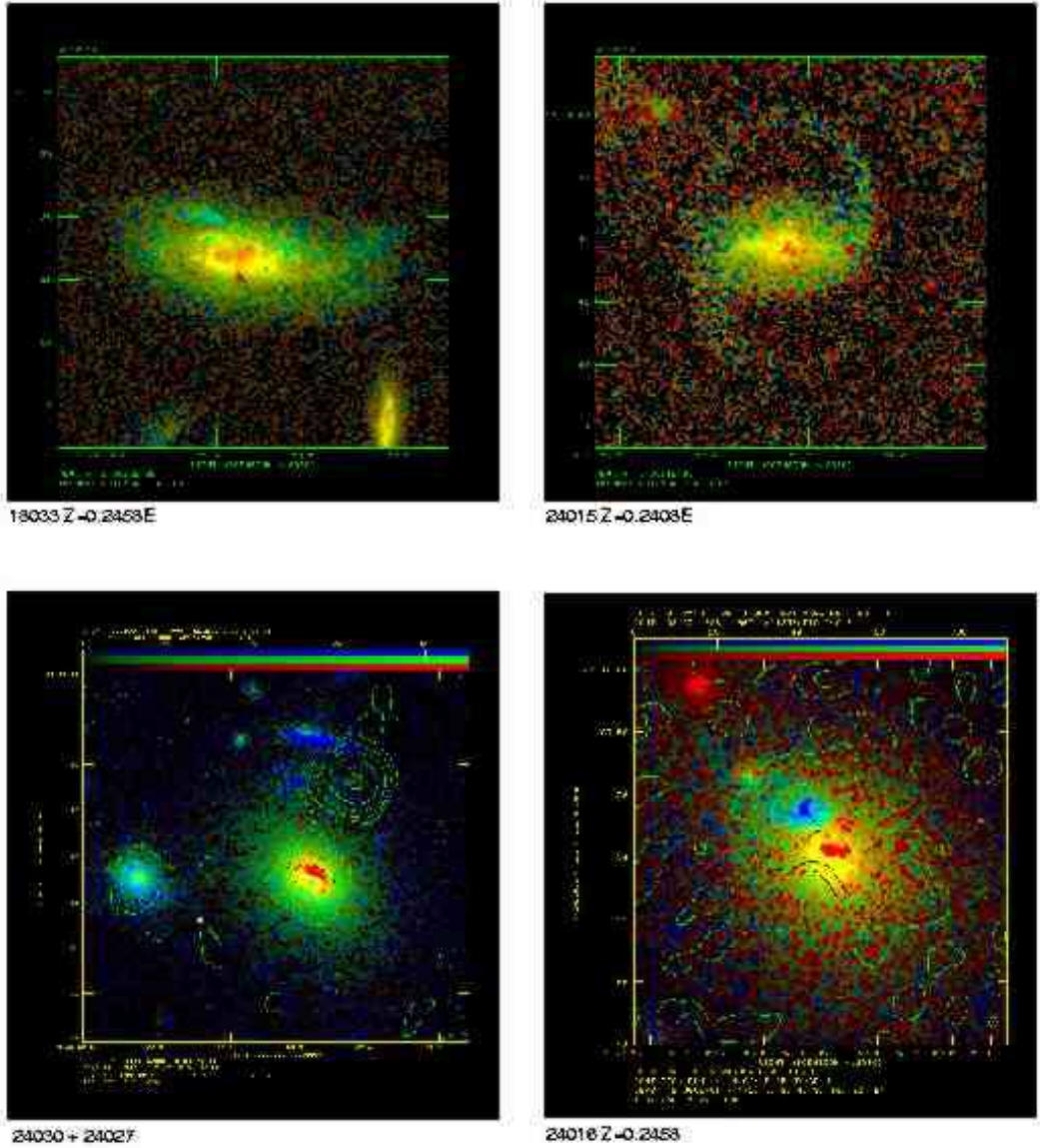}
\caption{Examples of the Diverse Structures seen HST Images of Radio
 Galaxies. The radio contours are overlaid in the lower two panels.
Comments on each image are given in the text.
  \label{HST}}
\end{figure}

	Figure~\ref{HST} shows images of
radio galaxies from the HST observations. The images are
true color in the sense that red represents redder light
and blue bluer light from our V and I images. In the bottom two 
cases, we overlay the contours of radio emission. 
	These galaxies give some insight into what the ground-based
observations are telling us about the population as a whole.

	Object 18033 has an SED which {\it hyperz} fits with an old Sb. The HST
morphology fits with this Hubble type. The galaxy has 
emission lines and aligned radio emission. It has a radio SFR of
about 6 M\subsun yr$^{-1}$. The size of the radio source is only
about 8 kpc (2 arcsec), about the size of the redder region
in the galaxy core, which may indicate dust obscuration of much of
the star-forming region. 

	Galaxy 24015 is best fit by an ``elliptical'' SED but
with an age of only 4.5 Gyr. However, the region fitted by
{\it hyperz} is interior to the the spiral arms seen with HST. 
The radio emission is aligned and consistent with a star-formation
rate of 2 M\subsun yr$^{-1}$. 

	Galaxy 24030 (on the east side of image) is also fit by a 
young ``elliptical'' SED but
shows all the other indications of star-formation at a modest rate.
A close look at the HST image suggests spiral structure. 

	Object 24027 (in the center of the image) is fit by a burst model of 
the same age as 24030 (4.5 Gyr). It shows
little evidence for star formation except for an $A_V$ of
0.6. The HST image has a red core also suggesting dust obscuration.
The radio source to the north and west of 24027 appears to be
unrelated since it is coincident with a very faint object on the
HST image; there is a faint {\it Chandra} source at that position.

	Object 24016 has an SED best fit by an old ``S0'' with some dust.
This is consistent with the red core and the lack of emission lines.
However, the radio emission appears quite extended and the NE lobe
of the radio source is coincident with a very blue object. This
blue object also shows strong [O II] emission in our narrow-band
image. Since this object is outside the region covered by the optical 
spectroscopy it is hard to guess what is going on here, whether
a superposition of an unrelated late-type galaxy or some sort
of jet-induced star formation. This main galaxy is still
most consistent with an old stellar population and AGN, but would require
more study to understand the pattern we see. 

	Overall, the HST imaging supports the conclusions drawn from
the ground-based observations. However, the galaxy types provided by 
{\it hyperz}
through the Bruzual-Charlot templates, need to be taken with a
grain of salt and only used as a general guide to the nature of
each galaxy in the region covered by the aperture used for the
photometry. Many of the objects do show morphological evidence for
dusty cores as well as from the SED fitting.

\subsection{Location of Activity in the Cluster}

\begin{figure}
\plotone{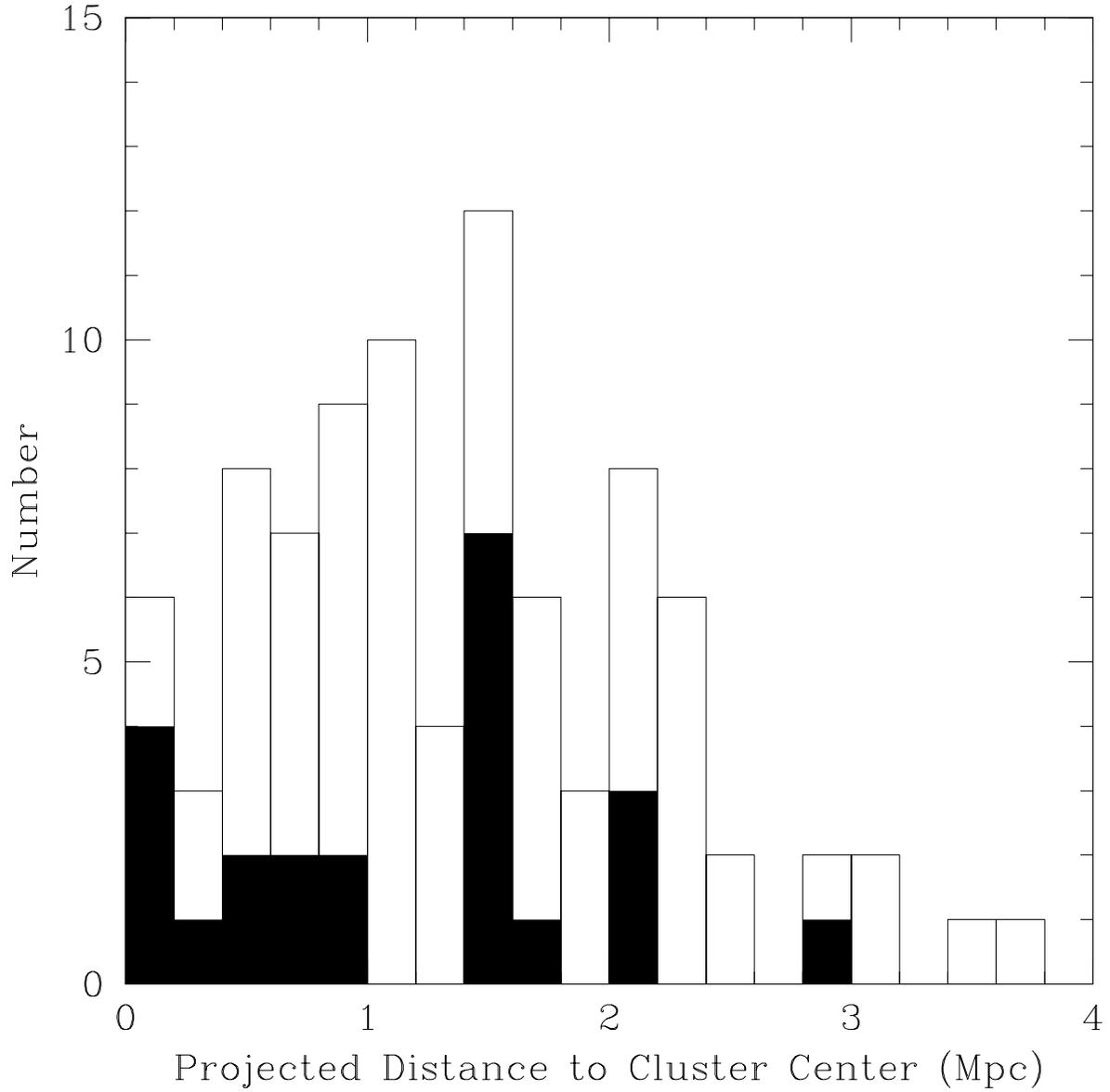}
\caption{Histogram of the distance of radio detected galaxies from
the cluster center in Mpc. The
solid boxes indicate objects, with
$SFI < 1$, the most likely AGN. \label{fig4}}
\end{figure}

	One of the important clues to understanding Abell 2125 is
the location of the radio galaxies with respect to the nominal
cluster core and the diffuse X-ray emission.
In figure~\ref{fig4} we show the histogram of the
projected distance from the nominal cluster center for the radio
cluster members. 
The black boxes indicate sources for which $SFI < 1$, the most
likely AGN. Clearly most of
the radio emitters are fairly
far from the cluster core, compared with the core radius from 
{\it Chandra} of $\sim 250$ kpc \citep{w03}.

    Three of the radio luminous objects are associated with the triple
system in the center of the cluster core and have SFI=0. Excluding
these objects, both the likely AGN and the entire sample have
median projected distances from the cluster core of $\sim 1.3$ Mpc.
Thus most of the radio detections, AGN and star-forming galaxies alike, lie far outside
the cluster core.

	\citet{w03} present the X-ray  results for imaging of the
diffuse structure in Abell 2125. The
diffuse X-ray structure of Abell 2125 gives a picture of a relatively cool,
moderately low luminosity system, $\sim 10^{44}$ erg s$^{-1}$. The
main cluster concentration has a temperature near 3 keV, while the
more diffuse structures to the SW have $T \sim 1$ keV. 
From \citet{w03} one can see that this extremely optically rich cluster 
(R=4) is not a particularly
impressive X-ray cluster. Its luminosity, temperature and density are
more like a much lower-richness cluster. Furthermore, the lower
surface brightness regions outside the bright core are
significantly cooler. The full size of the X-ray emitting region is at least
$4.5 \times 2.2$ Mpc in projection, centered SW of the core by about 1.2 Mpc.

\begin{figure}
\plotone{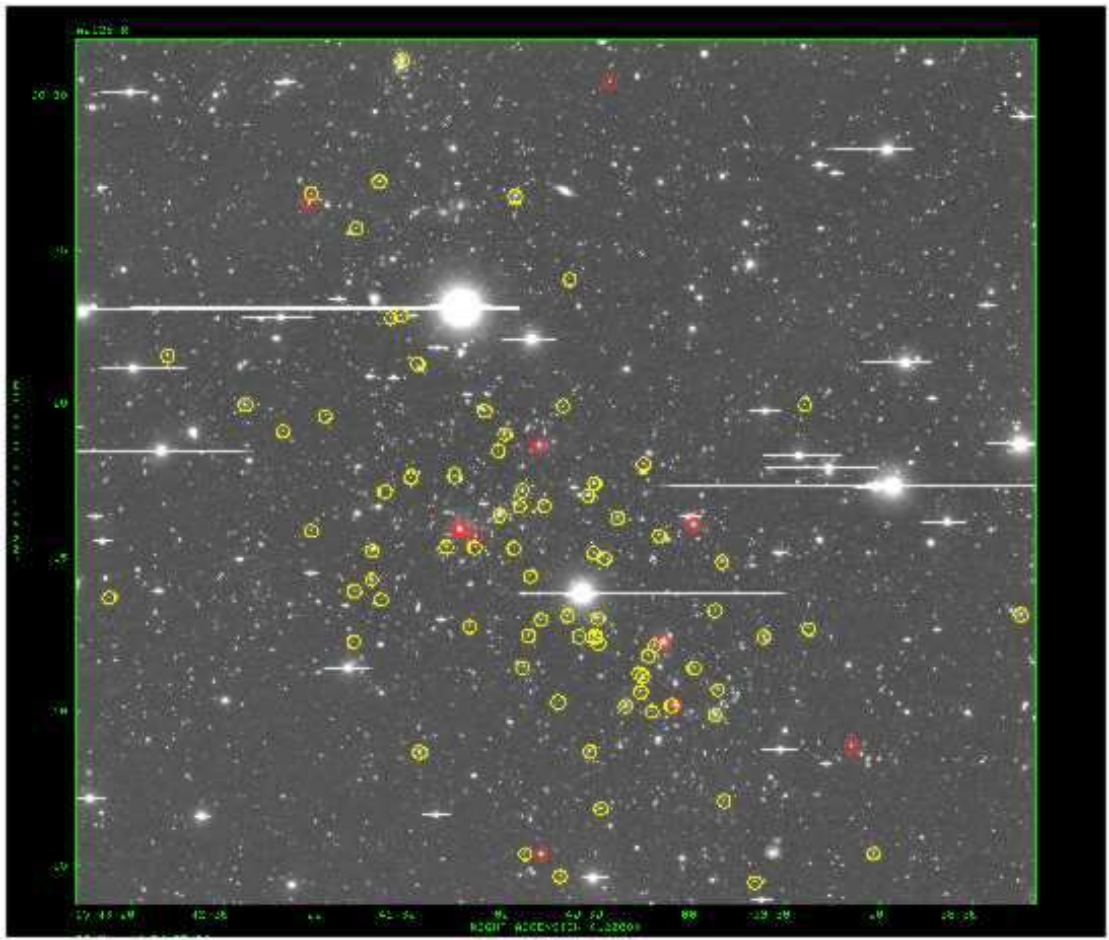}
\caption{Spectroscopically confirmed members with radio detections 
overlayed on the optical R image. Red circles are for radio sources 
with log $L_R > 22.7$. Yellow circles
are for $L_R \le 22.7$ The nominal cluster center is
at (15 41 14, 66 15 00); see figure~\ref{xor}. The field is
about 6.3 Mpc on a side in projection at Abell 2125. \label{rad}}
\end{figure}

\begin{figure}
\plotone{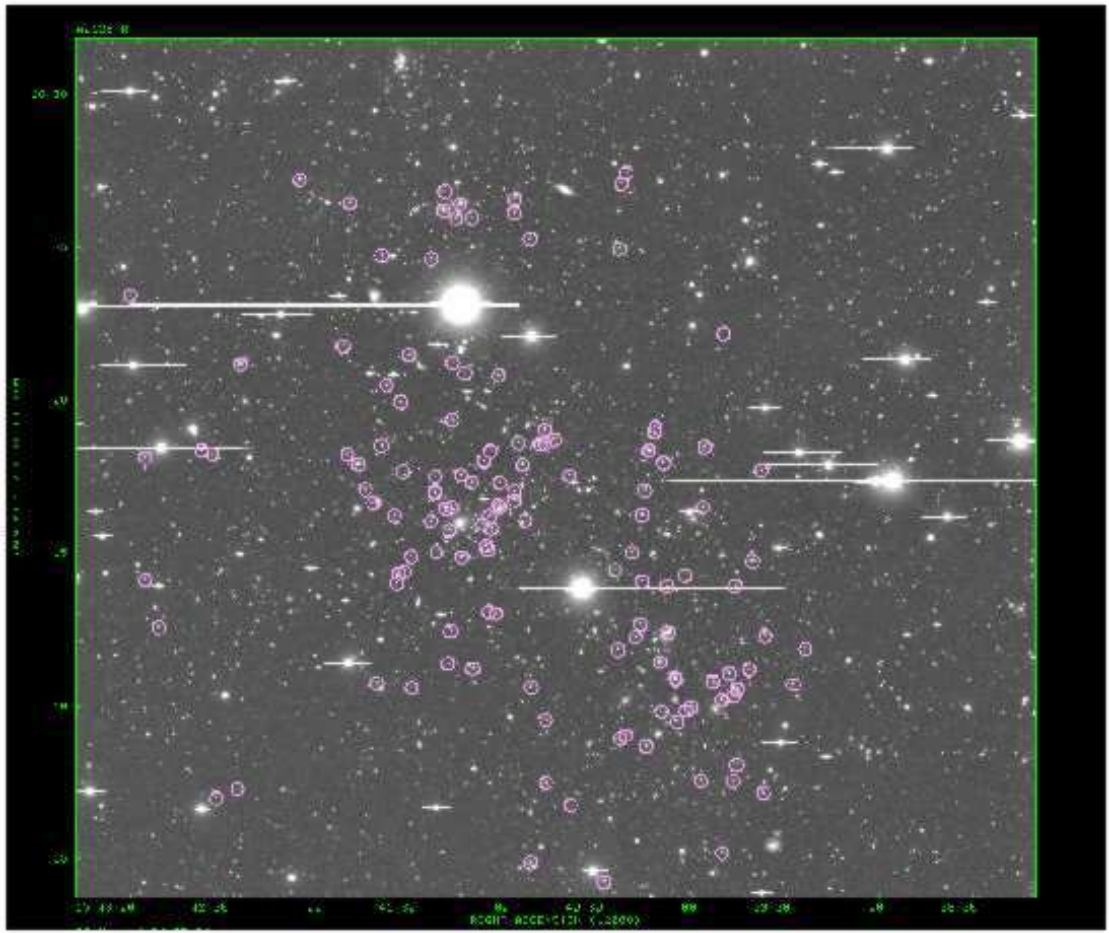}
\caption{Spectroscopically confirmed members without radio emission
overlayed on the R optical image. \label{nonrad}}
\end{figure}

	In figure~\ref{rad} we mark the radio emitters
on the optical field. Objects in red have $L_R > 22.7$,
clearly above the nominal break between AGNs
and starbursts \citep{m02}. In figure~\ref{nonrad} we show a similar
plot for the
cluster members without radio emission from \citet{m03}. Clearly,
the distribution of all the galaxies is centered about 1 Mpc southwest
of the nominal cluster core, near [15 41 15, 66 16 00] \citep{o99} 
and are distributed over a region about
$2\times6$ Mpc in projection. Unlike the impression of
two concentrations of radio emitters given by the smaller sample
in \citet{o99}, the lower luminosity radio emitters are
distributed almost uniformly over this large region, and appear even
somewhat less clustered than the non-radio emitters.  This pattern
suggests a fairly uniform distribution of radio-emitting
galaxies over the entire projected volume, perhaps avoiding the
core of this region. The nominal core of A2125 (where the
greatest projected local density of galaxies is found) is not
centered in the overall galaxy distribution, and the radio-emitting,
star-forming population seems not to be aware of this dense concentration of
galaxies. This situation is qualitatively similar to the Cygnus A
cluster in which the dense core containing the radio galaxy is offset from
the centroid of the overall galaxy distribution \citep{cyg}.
We find absolutely no correlation between distance from
the nominal cluster core and redshift for the radio population. The
radio galaxies  appear to have radial velocities centered on the
cluster mean but including the full range seen in Miller et al.(2003).
Thus there is no simple signature of an infalling radio population

\begin{figure}
\plotone{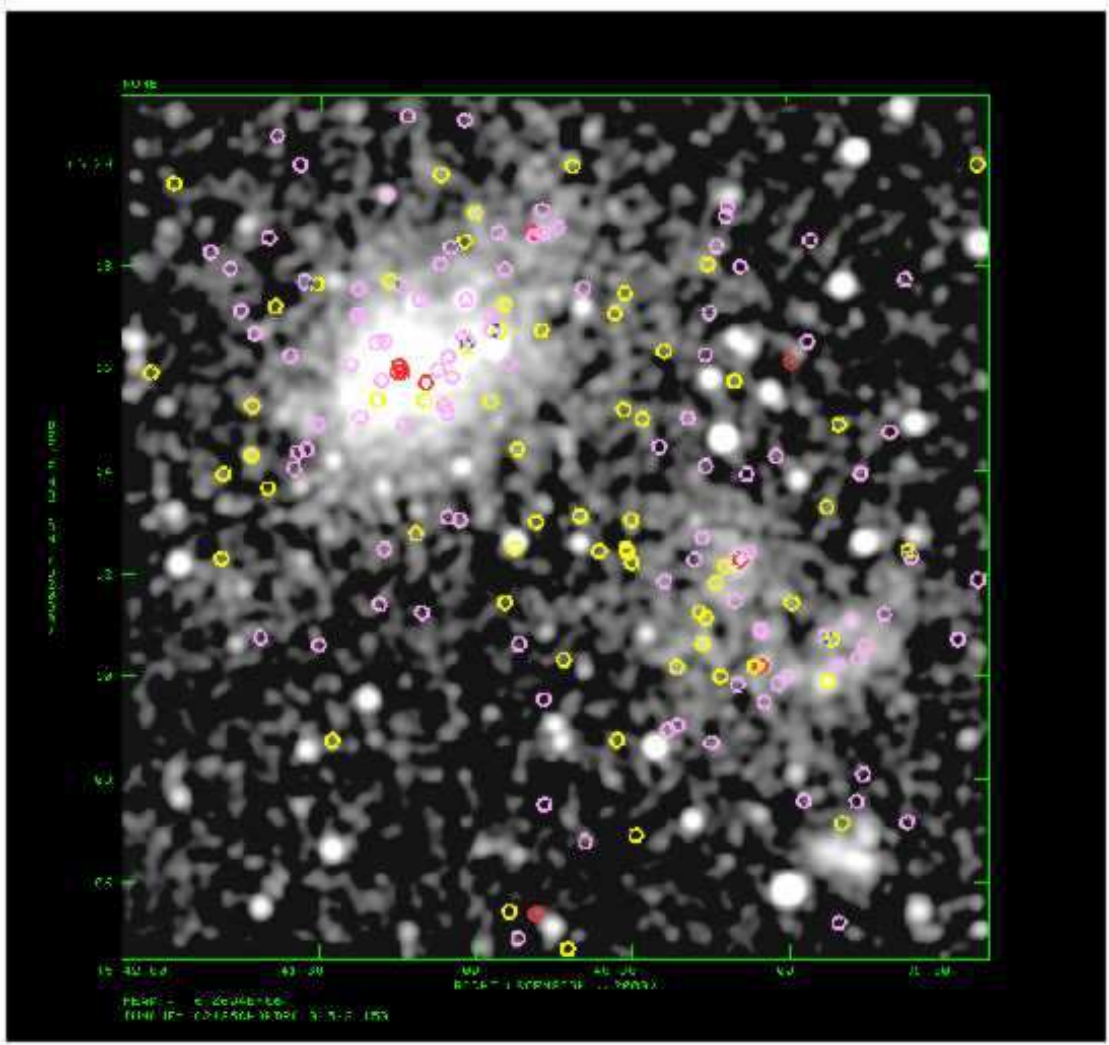}
\caption{Cluster Members overlaid on the the 0.5-2.0 keV {\it Chandra}
image of Abell 2125 convolved with a 15 arcsec Gaussian. The same
color coding is used as in figures \ref{rad} and \ref{nonrad}
\label{xor}}
\end{figure}

	In figure~\ref{xor}, we show the cluster members with the
same color coding as in figure~\ref{rad} and figure~\ref{nonrad}
overlaid on the the {\it Chandra} 0.5-2.0 keV image (as convolved with
a 15 arcsec circular Gaussian). The non-radio detected
galaxies do follow the the X-ray distribution in the cluster core and
to some extent the more diffuse emission to the southwest. However,
the low-luminosity radio emitters (mostly star-forming galaxies) do
not show much correlation with the X-ray detections except on the largest
scales.

\begin{figure}
\plotone{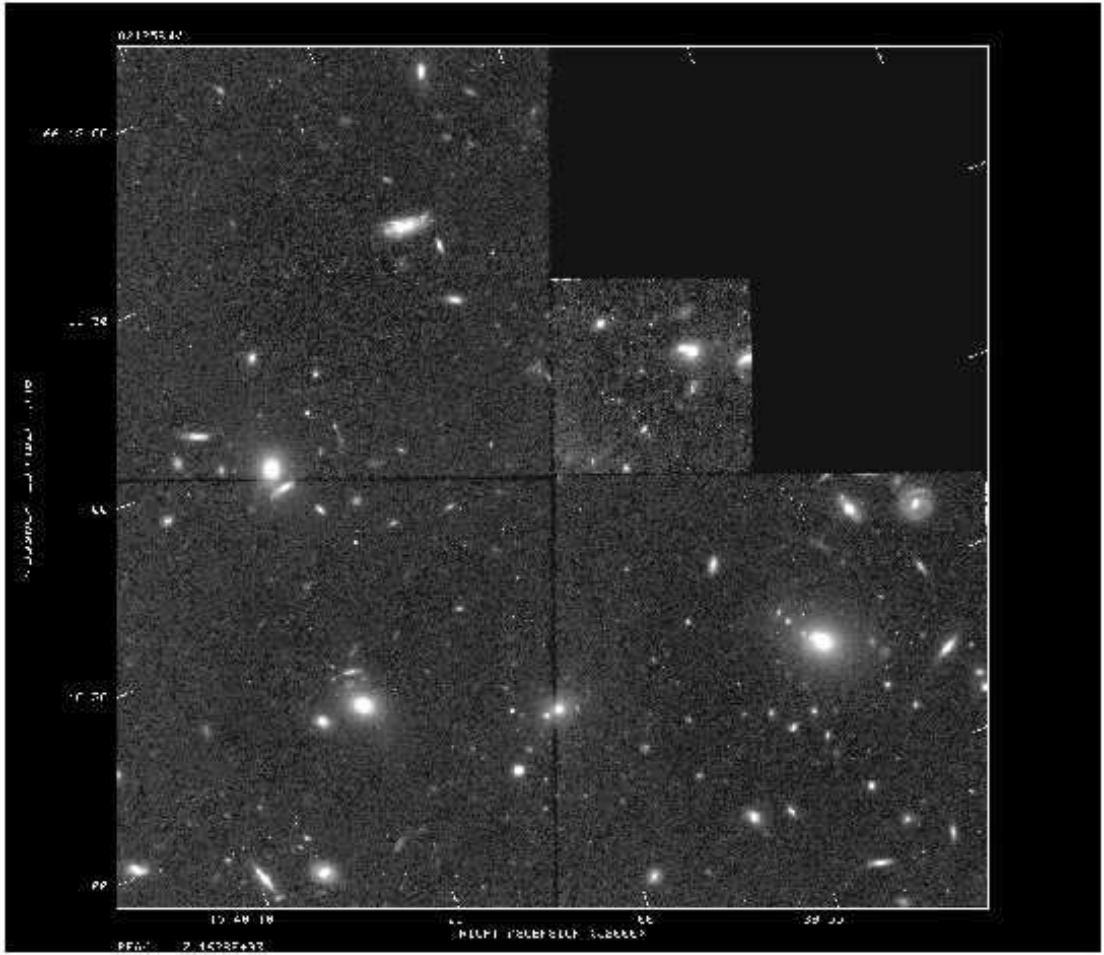}
\caption{HST WFPC2 Image of a region in the SW extension of
A2125 about 1.8 Mpc southwest of the nominal cluster center
in projection. The image is about 500 kpc on a side in
projection. Note the non-random distribution of galaxies consistent with
looking through a number of small groups instead of a random
cluster field.  \label{hstsw}}
\end{figure}

	Another perspective on Abell 2125 can seen in figure~\ref{hstsw},
the HST F814W image of one of the fields in a brighter part of the
SW X-ray concentration. Rather than a random distribution of galaxies
we find what appear to be several groups of a various sizes. A simple
2D substructure test, not taking into account the magnitudes of
the galaxies, shows substructure at $> 96$\% significance level
\citep{l00}. Thus in this part of the cluster, we seem to be looking 
through a system made up not of a
monolithic cluster but rather many small groups seen in projection.

	All of these results plus the broad distribution of
radio galaxies in figure~\ref{rad} suggest that we are dealing
with a complex interaction between at least two merging subsystems and
that much of the activity appears to be taking place on the outskirts
of the two systems, perhaps in group-like environments. The nominal
cluster core does not appear to be at the ``center'' of the large
scale cluster but is only one over-dense region in a much larger, complex
system. The X-ray luminosity and temperature are also much more
consistent with this picture than for a single, monolithic cluster.

\section{Discussion}

\subsection{General Picture of Abell 2125}

	From our optical and X-ray observations of Abell 2125 we conclude
that the vast majority of the radio emitting 
galaxies in the cluster reflect star formation.
The SFRs implied by this emission are relatively
modest, $1-10$ M\subsun yr$^{-1}$. Only a handful of the radio
emitters are compact X-ray AGN, about 10\%. Most of these are close to
our  {\it Chandra} detection limit. Surprisingly, at least 12 more
radio objects in the {\it Chandra}
field are not detected by {\it Chandra} but appear to be AGN, based
on a SFI $< 1$. Several of these are quite clearly FR I or FR II radio
galaxies \citep{dow}.
Essentially  all galaxies which are at least 30\% brighter than
$L_{*}$ appear to
be AGN driven, along with a much smaller percentage of the lower
optical-luminosity galaxies. Our 80 ksec {\it Chandra} exposure
does not seem to be deep enough to detect most radio AGN at $z=0.25$
found in our VLA radio survey of about the same total duration.  

	Except for four sources in the cluster core, which we
will discuss in another paper, both the AGN and the star-forming
objects are distributed on a scale of several Mpc and are less
clustered than the spectroscopically confirmed non-radio members.
Thus the increase in activity we are seeing in Abell 2125 with respect to
other rich clusters seems to be taking place on scales closer to
a supercluster than a typical cluster core.

	From figure~\ref{rad}, we see that the distribution of 
lower radio luminosity detections is offset from the nominal
cluster center about about 1 Mpc and distributed
rather uniformly over a region $2\times 6$ Mpc in projection, much
like the X-ray emission. In figure~\ref{nonrad}, we show the locations
of the known cluster members without radio detections. One can see
that these galaxies appear more clustered than the radio population.
Thus the projected distribution suggests that the low luminosity
radio population is less clustered than the radio-quiet population
and thus is likely to be located in lower density regions in the
cluster complex.

	From \citet{m03}, we see that the entire system can be modeled 
as an ongoing cluster-cluster merger seen at an angle to the
line-of-sight of about 30 degrees. In this picture, the excess star formation
is taking place on the scale of the merger, not in the cluster
core. Since the major axis of the merging system is near our
line-of-sight, the lower density outlying regions of the cluster
are projected closer to cluster center than for a more random
orientation. This makes the cluster appear richer and accounts
for the low X-ray luminosity for such an apparently rich cluster
\citep{m03}.
In fact, clusters near the upper end of Abell 
richness must tend to be the ones with their major axis in our line-of-site,
since more of the galaxies
will be projected into the 2D circle on the sky where richness is
estimated. 
Since the outer parts of clusters are bluer than the cores, such
clusters should tend to have larger blue fractions than average,
since more of the outer population are projected on the cluster
core where the Butcher-Oemler effect is measured \citep{bo}.
Thus projection effects explain some of the properties of Abell 2125.

	Examination of the outer parts of the cluster with HST
(figure~\ref{hstsw}) shows that many of the galaxies appear to be 
in small groups. This
is an environment one might expect to find on the outskirts of an
Abell cluster, in the moderate-density regions of a supercluster. Such environments are more
favorable to galaxy interactions and mergers than in regions
near the dense core of a cluster. Thus the broad distribution
of  excess star-forming radio galaxies is consistent with most of the
activity we see taking place in such medium density, group environments.

       The important question is whether projection effects explain
the entire radio excess in Abell 2125 relative to other similar
clusters. From \citet{MO03}, the median radio fraction for 
clusters of similar richness and redshift, considering
detected cluster members within 2.5 Mpc of the cluster center, is
about 2\%. For Abell 2125, the radio fraction is 9\%. 
For clusters similar to Abell 2125 in redshift and richness, only
$\sim 20$\% of the brighter galaxies used in the counts out to 2 Mpc
lie within 0.5 Mpc of the cluster core.  
Thus the richness and radio fraction estimate both describe primarily the
properties of galaxies far from the cluster core.  

       The well-known morphology-density relation \citep{d80} shows
that at the densities found in rich cluster cores, very few late type
galaxies are found and those show evidence for suppressed star formation
\citep{h98}. However, outside the cluster core at distances $\sim 1$
Mpc from the cluster center, the fraction of late type galaxies
typically rises to $\sim 40$\%, increasing only slowly
beyond that distance \citep{wg}.  Furthermore, \citet{h98} 
suggest that star formation is most active in such intermediate
density regions, where galaxy interactions should be most common. 

      It is likely we are seeing Abell 2125 in some sort of
projection. However, as long as most of the observed cluster
members are in intermediate
density regions, we would expect only a small bias to higher SFR
systems with respect to clusters whose major axis is nearer the plane
of the sky. However, the excess radio galaxy fraction 
is  four to five times higher than we typically see in  other similar
rich clusters. In order to explain the high radio fraction in Abell
2125 some other effect must be important. Such an effect could well be
that we see Abell 2125 at a special time. 
\citet{m03} argue that the dynamics
of Abell 2125 are consistent with such a picture. Their KMM
substructure analysis
of the positions and radial velocities of 224 galaxies associated with 
the cluster find that the
Abell 2125 is most consistent with two (or possibly three) velocity
systems seen in projection along the line-of-sight.  A comparison of
the observed dynamics
with the $n$-body merger simulations of \citet{prbb} show an excellent
match to a cluster-cluster merger seen at a viewing angle of about
30 degrees from the line-of-site and within 0.2 Gyr of core passage.
Assuming this physical situation, the  mass  of  the entire system
also lies within a reasonable range near $10^{15}$ M\subsun. 
Given this picture, several physical mechanisms could be
responsible for most of the increase in activity.

\subsection{Mechanisms}

	Numerous mechanisms have been considered which affect
the SFR of galaxies in clusters. Certainly, the rate of
galaxy mergers, ISM stripping,
galaxy harassment, changes in the external pressure and tidal
effects all have the potential to increase or decrease the SFR
of a galaxy. Most of the work on these subjects has considered
galaxies falling radially into a massive cluster. However, the
physical situation we are observing appears to be different, i.e.
two massive cluster and their associated lower density outskirts
involved in a major merger, probably near the point of core passage.
Since we are not aware of such a simulation designed to look for the 
effects relevant to star-formation we will need to extrapolate a bit
using the physical effects which have been discussed in a less extreme
context.

	In a high velocity cluster-cluster merger like Abell 2125, one 
would expect galaxy-galaxy mergers to be disfavored, since the probability 
of a galaxy-galaxy merger is proportional to the inverse fourth power
of the local velocity dispersion \citep{gn03a}. However, the 
details of group-group mergers in such a system might complicate this
argument. Nonetheless, we see no direct evidence that galaxy-galaxy
mergers play a major role in Abell 2125.

	\citet{fn99} conclude that ram pressure on a galaxy
radially infalling into a cluster can increase the SFR by a factor
of 2 near the cluster core before the rate decreases due to the
stripping of the gas. In a complex cluster-cluster merger, the   
velocity of galaxies relative to the local medium could be much larger
than in the infall case. Thus the total ram pressure could easily
rise above the values found in the center of an isolated cluster
Furthermore, some of the systems in figure~\ref{RO2} show evidence
of non-aligned radio emission which could indicate stripping,
lending support to ram pressure as a contributer to the increased
SFR. 

	Galaxy harassment \citep{mlk98}, involves the cumulative
effects of fast encounters between galaxies in clusters. Most
of the impact of this mechanism involves lower-mass galaxies near
the cluster core. Since most
of the activity we see is far from the core of Abell 2125 and involves
relatively massive galaxies, it seem unlikely this mechanism has
much to do with the effects we are seeing.

	Tidal effects in clusters have been recently considered by 
a number of papers as a way to affect galaxy morphology and SFRs.
\citet{bv} and \citet{hb} argue that the tidal effect of the cluster 
potential on an infalling galaxy would create more star-formation
activity than stripping by the IGM. \citet{b99} studied the case of
the tidal effects on a spiral in a group, as the group falls into a 
more massive cluster; besides driving a
transient starburst, the variable tidal field of the group heats the
disk and ultimately leads to an S0 morphology. \citet{gn03a} simulates the
general case of galaxies falling into a cluster in the presence of 
the time varying tidal field resulting from other substructure which also is
in the process of falling in. The simulation shows that in the frame
of an infalling galaxy there exists a strongly time-variable tidal 
field with multiple major events far from the cluster
core. \citet{gn03b} concludes that this process leads to the
transformation of spirals into S0's. 
 
\subsection{Application to Abell 2125}

	The case we see in Abell 2125 is somewhat different than
discussed in the previous section in that we seem to be seeing
a major merger close to core passage. Thus the mechanisms discussed
above need to be considered at this special time, which is supported by the fact that
Abell 2125 stands out from other rich clusters that have been studied at
similar redshifts \citep{MO03}. At a later time one should expect
A2125 to have settled down and to have evolved into a rich cluster like
those we see locally.

	In such a dynamic case, the time-variable tidal effects
must be especially extreme. It thus seems likely that the effects
described by \citet{b99} and \citet{gn03a} must be especially
important. However, not enough time will have passed for the systems
to reach their ultimate fate. The observations show an excess of
galaxies with active star-formation but 37\% of these
systems have optical concentration indices
lying between disks and $r^{1/4}$
laws. They also have SEDs far from young objects one might associate
with bursts of recent star-formation. 

	This pattern seems consistent with star formation related to
the strongly time-variable tidal forces  
as analyzed in \citet{b99} and \citet{gn03a}. Once the event is over, these
galaxies seem on their way to becoming dull S0's as described 
by \citet{gn03a}. The distribution of concentration indices suggests
that many of our detected systems are neither pure disks nor 
$r^{1/4}$ spheroids. Thus the structures of many of the radio galaxies
(see figure~\ref{CI}) are
consistent with objects which have experienced tidal heating. 
However, the analysis of \citet{b99} and \citet{gn03b} suggest that the
timescale for the transition from disk to S0 is very long, many Gyrs,
while the gas-dynamic processes which led to star-formation have much
shorter timescales. Thus the broad distribution of concentration
indices associated with the radio galaxies either suggests that tidal
heating has been taking place for some time or that the tidal forces
are larger in the Abell 2125 case leading to more rapid evolution of
the stellar distribution. A more targeted analysis of tidal effects
during a major cluster-cluster merger is needed to understand better
how tidal (and other) effects would affect a system like Abell 2125.
But given that cluster-cluster mergers are quite common events, our
observational results suggest that core passage during a major 
cluster-cluster is may be an especially important time in galaxy evolution.

\section{Conclusions}

	The excess levels of radio emission in Abell 2125 are dominantly
related to star formation. Abell 2125 is a cluster-cluster
merger seen in projection. Active star formation is not associated
with the cluster core but spread fairly uniformly throughout
a $2\times 6$ Mpc region in projection. This activity
must be taking place in intermediate density regions, mostly in groups
associated with the outskirts of the merging clusters. 

	The cluster-cluster merger seen in projection accounts for
the relatively low X-ray luminosity and temperature, combined with
the very high richness and relatively high Butcher-Oemler blue
fraction. The excess radio population also is at least partly
explained by projection effects. However, the large radio population
seems also to require some other mechanism to explain the four to five
times larger radio fraction than is seen in other such clusters.
This extra factor may be due to the cluster-cluster merger
being seen very close to core passage where the interactions of
the two cluster systems should be greatest.

It seems likely that star-formation has been stimulated in the
outlying group environments associated with the clusters during the  
time near the core passage. The variations in the tidal field,
experienced by each galaxy due to the changing substructure and
its position relative to each galaxy seems to be the most likely
origin for the enhanced SFRs seen in Abell 2125. The activity in Abell
2125,
combined with its dynamical state close to core passage of a
major cluster-cluster merger, suggests that this phase in cluster
evolution may be particularly important for galaxy evolution.
	
	The authors thank A. Klypin  for useful discussions about
the large scale structure and J. Eilek for comments on the text.

\clearpage
\clearpage
\clearpage
\clearpage
\clearpage
\clearpage

\begin{deluxetable}{lrrrrrll}
\tablecolumns{8}
\tablewidth{0pt}
\tablecaption{A2125 Member Radio Sources\label{SL}}
\tablenum{1}
\pagestyle{empty}
\tablehead{
\colhead{Name} & 
\colhead{RA(2000.0)} &
\colhead{Dec(2000.0)} &
\colhead{Dist} &
\colhead{Peak} &
\colhead{Total} &
\colhead{Size} &
\colhead{z}\\
\colhead{} &
\colhead{} &
\colhead{} &
\colhead{Mpc} &
\colhead{$\mu$Jy/b} &
\colhead{$\mu$Jy} &
\colhead{arcsec} &
\colhead{}}  
\startdata
17017&15 38 13.80(0.08)& 66 13 05.9(0.5)& 18.3&   32.1&  148.5(  33.)&$<$2.2         &0.2475 \\
23009&15 39 01.94(0.09)& 66 05 24.2(0.5)& 16.5&   39.8&  300.6(  85.)&   4.1         &0.2575 \\
23016&15 39 08.69(0.02)& 66 08 52.8(0.1)& 14.1&  764.1&60235.0(1829.)&  25.0         &0.2449 \\
18013&15 39 22.18(0.05)& 66 12 41.6(0.3)& 11.5&   75.1&  152.4(  15.)&$<$1.3         &0.2449e\\
11700&15 39 23.16(0.09)& 66 19 57.7(0.5)& 12.2&   25.3&   53.3(  15.)&$<$2.5         &0.2442e\\
18023&15 39 36.86(0.04)& 66 12 27.0(0.3)& 10.1&  109.1&  191.6(  14.)&$<$1.2         &0.2448 \\
24007&15 39 40.10(0.05)& 66 04 28.7(0.3)& 14.2&  128.5&  364.1(  26.)&   1.0x0p=46   &0.2468e\\
24012&15 39 49.73(0.06)& 66 07 99.1(0.4)& 11.6&   62.2&  170.8(  26.)&   1.6x0.5p=5  &0.2445e\\
18700&15 39 50.14(0.08)& 66 14 54.7(0.5)&  8.4&   27.1&  39.7(   10.)&$<$1.8         &0.2458 \\
24015&15 39 51.62(0.08)& 66 10 43.8(0.5)&  9.3&   42.6&  114.7(  29.)&   2.6x0p=108  &0.2408e\\
18030&15 39 52.47(0.06)& 66 13 18.5(0.4)&  8.4&   55.6&   83.3(  11.)&$<$0.9         &0.2482e\\
24016&15 39 52.52(0.07)& 66 09 54.3(0.4)&  9.7&   50.6&  270.1(  48.)&   3.6         &0.2458 \\
18033&15 39 59.31(0.04)& 66 11 26.6(0.3)&  8.3&  166.4&  348.4(  23.)&   2.0x0.7p=71 &0.2458e\\
18072&15 39 59.43(0.02)& 66 16 07.5(0.1)&  7.6&  600.4&10668.0( 335.)&  99.0         &0.2459 \\
24027&15 40 05.34(0.03)& 66 10 12.9(0.2)&  8.4&  359.5&  970.0(  45.)&   4.6         &0.2425 \\
24030&15 40 06.60(0.08)& 66 10 12.8(0.5)&  8.3&   38.3&   82.6(  21.)&   1.9x0p=156  &0.2436e\\
18041&15 40 09.11(0.05)& 66 12 17.1(0.3)&  7.1&  114.3&  390.0(  30.)&   2.3x0.6p=75 &0.2455 \\
18701&15 40 10.34(0.09)& 66 15 46.2(0.6)&  6.5&   32.3&  149.1(  46.)&   8.5x2.0p=156 &0.2477 \\
18042&15 40 12.05(0.03)& 66 12 09.9(0.2)&  6.9&  167.5&  330.5(  15.)&$<$0.7         &0.2558 \\
24111&15 40 13.05(0.10)& 66 10 01.1(0.6)&  7.9&   27.3&  107.2(  42.)&   5.5x0p=93   &0.2460e\\
18045&15 40 13.98(0.09)& 66 11 50.7(0.5)&  6.8&   35.1&   62.1(  17.)&   1.6x0p=71   &0.2458e\\
18047&15 40 15.38(0.08)& 66 18 02.5(0.5)&  6.6&   43.6&  101.2(  22.)&   2.2x0.6p=54 &0.2433e\\
18048&15 40 15.84(0.04)& 66 11 09.9(0.3)&  7.0&  134.6&  334.3(  23.)&   1.7x1.5p=57 &0.2455e\\
24701&15 40 16.49(0.08)& 66 10 39.7(0.5)&  7.2&   29.8&   39.9(   9.)&$<$1.9         &0.2559 \\
18702&15 40 17.14(0.12)& 66 11 15.9(0.7)&  6.8&   25.8&   37.6(  19.)&   2.7x0p=115  &0.2443e\\
24033&15 40 21.42(0.07)& 66 10 12.1(0.4)&  7.2&   36.6&   48.8(   9.)&$<$1.9         &0.2565 \\
18703&15 40 23.87(0.12)& 66 16 21.4(0.7)&  5.2&   21.8&   33.5(  18.)&   2.5x0p=84   &0.2500 \\
05012&15 40 26.03(0.06)& 66 30 29.5(0.3)& 16.2&   88.9&  426.7(  49.)&   1.8x0.8p155 &0.2573 \\
18070&15 40 28.02(0.09)& 66 15 01.9(0.5)&  4.6&   29.8&   50.3(  15.)&   1.5x0.6p=80 &0.2475e\\
24115&15 40 29.36(0.07)& 66 06 55.1(0.4)&  9.3&   35.6&   56.6(  11.)&$<$1.3         &0.2421 \\
18056&15 40 30.13(0.04)& 66 12 14.2(0.3)&  5.2&  144.3&  195.0(  12.)&   0.9x0p=123  &0.2456e\\
18057&15 40 30.23(0.05)& 66 13 04.1(0.3)&  4.8&  114.7&  151.2(  12.)&   1.20x0p=27  &0.2494e\\
18300&15 40 30.98(0.09)& 66 12 26.2(0.5)&  5.0&   25.3&   29.2(   8.)&$<$2.0         &0.2458e\\
18060&15 40 31.21(0.06)& 66 12 30.5(0.3)&  5.0&   81.2&  199.2(  24.)&   2.6x1.5p=3   &0.2474e\\
18061&15 40 31.27(0.08)& 66 17 28.9(0.5)&  5.0&   43.9&   98.2(  24.)&   3.1x1.9p=4  &0.2430 \\
18062&15 40 31.55(0.06)& 66 15 12.6(0.3)&  4.3&   76.1&  149.3(  18.)&   1.9x0.8p=66 &0.2444e\\
25001&15 40 32.93(0.09)& 66 08 46.7(0.5)&  7.5&   35.6&  143.4(  41.)&   4.9x2.5p=68 &0.2430e\\
00002&15 40 33.28(0.04)& 66 17 05.8(0.2)&  4.6&  137.4&  156.7(   9.)&$<$0.9         &0.2548 \\
00600&15 40 36.31(0.07)& 66 12 28.0(0.4)&  4.6&   35.0&   39.4(   8.)&$<$1.5         &0.2423e\\
12004&15 40 39.09(0.09)& 66 24 05.8(0.5)&  9.7&   40.0&   67.9(  19.)&   1.7x0p=15   &0.2381e\\
00007&15 40 40.05(0.07)& 66 13 08.9(0.4)&  3.9&   56.3&  150.4(  24.)&   2.7x1.1p=104 &0.2462e\\
12007&15 40 41.44(0.09)& 66 19 57.2(0.6)&  5.9&   39.0&   74.8(  23.)&   3.1x0p=137  &0.2429e\\
25100&15 40 42.46(0.09)& 66 04 42.5(0.5)& 10.8&   34.9&  210.1(  57.)&   4.8x2.3p=107 &0.2521e\\
25010&15 40 43.17(0.03)& 66 10 20.6(0.2)&  5.6&  222.0&  264.7(  12.)&$<$0.7         &0.2558 \\
00700&15 40 47.55(0.10)& 66 16 45.5(0.6)&  3.2&   24.8&   40.6(  16.)&   2.1x0p=47   &0.2469 \\
00014&15 40 48.54(0.08)& 66 13 01.3(0.5)&  3.2&   34.5&  182.2(  42.)&   5.5x4.7p=127 &0.2423 \\
25702&15 40 48.63(0.06)& 66 05 24.1(0.4)&  9.9&   79.6&  419.8(  52.)&  11.03x6.5p=14 &0.2561 \\
00104&15 40 49.16(0.02)& 66 18 39.8(0.1)&  4.4&  504.0&16108.0( 497.)&  25.0         &0.2433 \\
00020&15 40 52.14(0.08)& 66 14 27.0(0.5)&  2.3&   32.1&  144.6(  37.)&   4.5x1.7p=9  &0.2419e\\
00022&15 40 52.63(0.05)& 66 12 30.4(0.3)&  3.3&   97.2&  190.5(  16.)&   1.5x1.3p=96 &0.2588e\\
25015&15 40 53.63(0.05)& 66 05 26.4(0.3)&  9.8&   94.3&  153.5(  12.)&$<$1.0         &0.2508e\\
00026&15 40 54.49(0.04)& 66 11 27.4(0.2)&  4.1&  191.3&  210.0(  10.)&$<$0.7         &0.2484e\\
00027&15 40 54.67(0.03)& 66 17 15.7(0.2)&  3.0&  246.0&  310.1(  13.)&   0.8x0.5p=119 &0.2560e\\
00028&15 40 55.56(0.06)& 66 16 44.7(0.4)&  2.5&   53.3&   55.5(   7.)&$<$1.1         &0.2486 \\
12027&15 40 56.88(0.06)& 66 26 45.6(0.3)& 11.9&  122.5&  308.6(  35.)&   3.1x0.6p=167 &0.2463 \\
00031&15 40 57.30(0.08)& 66 15 22.5(0.5)&  1.7&   51.6&   96.0(  19.)&   2.7x0p=16   &0.2545e\\
00034&15 41 00.33(0.07)& 66 19 02.5(0.4)&  4.3&   37.8&   41.9(   8.)&$<$1.0         &0.2395 \\
00039&15 41 01.93(0.03)& 66 16 26.6(0.2)&  1.9&  202.7&  207.2(   9.)&$<$0.8         &0.2457 \\
00201&15 41 02.33(0.06)& 66 18 29.4(0.4)&  3.7&   38.7&   72.7(  10.)&$<$5.4         &0.2454e\\
12033&15 41 06.84(0.09)& 66 19 47.5(0.5)&  4.8&   32.0&  125.0(  35.)&   4.2x1.7p=55 &0.2406e\\
00047&15 41 09.73(0.02)& 66 15 44.5(0.1)&  0.9&21783.0&22944.0( 689.)&  14.0         &0.2528e\\
00703&15 41 09.97(0.08)& 66 15 23.6(0.5)&  0.6&   35.6&  156.2(  38.)&   6.3x4.7p=55 &0.2494 \\
00051&15 41 11.59(0.09)& 66 12 47.9(0.6)&  2.2&   33.0&   54.4(  18.)&   2.5x0p=113  &0.2473e\\
00057&15 41 14.37(0.02)& 66 15 57.1(0.1)&  1.0& 1300.0& 2910.0(  90.)&   2.8x0.8p=12 &0.2518 \\
00105&15 41 14.87(0.04)& 66 16 03.8(0.2)&  1.1&  272.0&  677.0(  33.)&   4.0         &0.2470 \\
00106&15 41 15.24(0.02)& 66 15 56.7(0.1)&  1.0& 1052.0&15712.0( 477.)&  23.0         &0.2466 \\
00060&15 41 16.60(0.08)& 66 17 42.6(0.5)&  2.7&   47.4&  112.1(  23.)&   2.9x0.8p=178 &0.2422e\\
00701&15 41 19.06(0.09)& 66 15 24.2(0.5)&  0.6&   24.2&   24.2(   7.)&$<$2.3         &0.2480 \\
25701&15 41 27.59(0.09)& 66  8 45.6(0.5)&  6.4&   25.6&   32.5(   9.)&$<$1.3         &0.2454 \\
12068&15 41 28.51(0.06)& 66 21 20.1(0.4)&  6.5&   53.4&   67.1(   9.)&$<$1.4            &0.2484e\\
00077&15 41 30.76(0.06)& 66 17 39.5(0.3)&  3.1&   76.5&  132.4(  15.)&   1.6x0.9p=163 &0.2516e\\
06008&15 41 33.87(0.05)& 66 31 11.7(0.3)& 16.3&   96.4&  347.1(  27.)&$<$1.2         &0.2369 \\
12058&15 41 33.89(0.05)& 66 22 54.7(0.3)&  8.2&  105.5&  149.6(  11.)&$<$0.9         &0.2514 \\
12100&15 41 37.29(0.09)& 66 22 49.8(0.5)&  8.2&   36.6&  104.4(  29.)&   2.9x0p=104  &0.2488e\\
00084&15 41 38.90(0.06)& 66 17 13.3(0.4)&  3.3&   62.6&   82.7(  11.)&   1.2x0.2p=24 &0.2397e\\
00085&15 41 40.16(0.08)& 66 13 40.3(0.5)&  3.0&   41.0&  145.8(  31.)&   3.7x1.4p=37 &0.2517 \\
12702&15 41 41.13(0.08)& 66 27 12.7(0.5)& 12.5&   29.3&   65.3(  16.)&$<$3.1         &0.2398e\\
00089&15 41 43.23(0.05)& 66 15 16.6(0.3)&  3.0&  128.8&  152.1(  12.)&   1.4x0.4p=145 &0.2518 \\
00704&15 41 43.28(0.06)& 66 14 19.2(0.3)&  3.0&   97.2&  297.8(  34.)&   8.1x6.1p=179 &0.2547 \\
12061&15 41 48.61(0.05)& 66 25 45.4(0.3)& 11.3&   80.0&  151.1(  14.)&$<$0.9         &0.2492e\\
00705&15 41 48.77(0.07)& 66 13 57.2(0.4)&  3.7&   62.8&  193.8(  34.)&   9.3x5.0p=129 &0.2444 \\
00098&15 41 49.03(0.06)& 66 12 18.2(0.3)&  4.4&   66.1&   74.6(   8.)&$<$0.9         &0.2473e\\
13700&15 41 58.49(0.08)& 66 19 36.4(0.5)&  6.4&   27.7&   35.0(   9.)&$<$1.4         &0.2534 \\
19011&15 42 02.85(0.10)& 66 15 54.8(0.6)&  5.0&   38.2&   69.4(  25.)&   4.6x0p=101  &0.2412e\\
13003&15 42 03.39(0.08)& 66 26 48.7(0.5)& 12.8&   45.3&  179.4(  43.)&   2.8x0p=101  &0.2468e\\
13004&15 42 03.82(0.06)& 66 26 31.8(0.4)& 12.6&   80.4&  458.6(  60.)&   4.3x0p=151  &0.2458e\\
13008&15 42 12.07(0.09)& 66 19 07.0(0.6)&  7.1&   34.0&   89.5(  28.)&   3.2x0p=121  &0.2501e\\
13015&15 42 24.27(0.05)& 66 19 58.3(0.3)&  8.6&  141.7&  231.2(  17.)&   1.3x0p=127  &0.2473 \\
13702&15 42 49.60(0.07)& 66 21 33.2(0.4)& 11.6&   71.5&  224.4(  40.)&   6.6x4.1p=175 &0.2390\\
19061&15 43 07.80(0.08)& 66 13 41.4(0.5)& 11.5&   46.5&  129.0(  29.)&   2.0x0p=72   &0.2394e\\
\enddata
\end{deluxetable}

\clearpage

\begin{deluxetable}{cclc}
\tablewidth{0pt}
\tablecolumns{4}
\tablecaption{Templates\label{T}}
\tablenum{2}
\tablehead{
\colhead{Spec Type} & 
\colhead{Number} &
\colhead{SFR} &
\colhead{Timescale}}
\startdata
Burst&1&Single Burst& --- \\
E&2&Exponential&$\tau = 1$ Gyr\\
S0&3&Exponential&$\tau = 2$ Gyr\\
Sa&4&Exponential&$\tau = 3$ Gyr\\
Sb&5&Exponential&$\tau = 5$ Gyr\\
Sc&6&Exponential&$\tau = 15$ Gyr\\
Sd&7&Exponential&$\tau = 30$ Gyr\\
Im&8&Constant& --- \\
\enddata
\end{deluxetable}

\clearpage

\begin{deluxetable}{rrrrrrrrrrr}
\tablecolumns{11}
\tablewidth{0pt}
\tablecaption{A2125 Gunn-Oke Magnitudes\label{data}}
\tablenum{3}
\pagestyle{empty}
\tablehead{
\colhead{Name} & 
\colhead{U} &
\colhead{B} &
\colhead{V} &
\colhead{R} &
\colhead{I} &
\colhead{8010} &
\colhead{9170} &
\colhead{J} &
\colhead{H} &
\colhead{K}}  
\startdata
17017&22.08&20.65&19.36&18.45&17.93&17.94&17.50&$-9.00$&$-9.00$&$-9.00$\\
23009&22.71&21.38&20.08&19.10&18.47&18.44&18.05&$-9.00$&$-9.00$&$-9.00$\\
23016&22.30&21.00&19.74&18.84&18.39&18.35&17.99&$-9.00$&$-9.00$&$-9.00$\\
18013&23.14&21.95&20.92&20.12&19.71&19.73&19.33&$-9.00$&$-9.00$&$-9.00$\\
11700&22.99&21.84&20.81&19.98&19.53&19.57&19.20&$-9.00$&$-9.00$&$-9.00$\\
18023&22.39&21.01&19.72&18.76&18.19&18.22&17.78&$-9.00$&$-9.00$&$-9.00$\\
24007&22.70&21.57&20.57&19.66&19.10&19.17&18.88&$-9.00$&$-9.00$&$-9.00$\\
24012&22.33&21.29&20.21&19.30&18.79&18.74&18.39&17.94&17.50&17.36\\
18700&21.89&20.51&19.15&18.23&17.73&17.73&17.32&16.83&16.60&16.50\\
24015&21.77&20.99&20.05&19.32&18.91&18.86&18.61&18.24&17.97&17.92\\
18030&23.43&22.00&20.83&19.92&19.47&19.48&19.08&18.51&18.20&18.04\\
24016&21.08&20.01&18.87&17.97&17.47&17.47&17.11&16.62&16.42&16.32\\
18033&20.94&20.31&19.55&18.90&18.49&18.54&18.22&17.74&17.54&17.33\\
18072&21.91&20.43&19.13&18.19&17.65&17.75&17.22&16.81&16.61&16.51\\
24027&21.99&20.54&19.18&18.22&17.66&17.69&17.26&16.70&16.42&16.29\\
24030&22.24&21.48&20.59&19.89&19.50&19.49&19.18&19.03&18.73&18.76\\
18041&21.64&20.12&18.75&17.78&17.26&17.31&16.89&16.41&16.23&16.07\\
18071&21.92&20.75&19.60&18.75&18.26&18.35&17.90&17.33&17.01&16.95\\
18042&22.02&20.54&19.21&18.24&17.75&17.80&17.39&16.85&16.63&16.49\\
24111&21.27&20.74&20.10&19.61&19.52&19.56&19.22&18.93&18.82&18.67\\
18045&20.85&20.41&19.86&19.44&19.34&19.40&19.18&18.94&18.77&18.93\\
18047&21.15&20.64&20.01&19.51&19.29&19.36&19.10&18.80&18.72&18.63\\
18048&20.73&20.10&19.37&18.78&18.39&18.44&18.13&17.74&17.49&17.44\\
24701&22.33&20.99&19.66&18.70&18.24&18.26&17.91&17.26&17.08&16.96\\
18702&21.05&20.41&19.67&19.03&18.71&18.72&18.47&18.13&18.01&17.97\\
24033&22.04&20.61&19.27&18.31&17.79&17.82&17.47&16.87&16.61&16.50\\
18703&22.25&20.88&19.60&18.69&18.20&18.24&17.88&17.29&17.05&17.00\\
05012&21.96&20.77&19.68&18.81&18.30&18.32&17.87&$-9.00$&$-9.00$&$-9.00$\\
18070&22.23&21.70&21.06&20.48&20.39&20.42&20.10&19.88&19.56&19.47\\
24115&22.47&21.00&19.73&18.80&18.35&18.32&17.97&17.56&17.36&17.20\\
18056&21.90&21.02&20.03&19.25&18.80&18.88&18.48&17.92&17.75&17.45\\
18057&21.85&20.55&19.28&18.34&17.81&17.89&17.41&16.96&16.72&16.58\\
18300&21.68&21.10&20.29&19.76&19.51&19.65&19.45&18.96&18.82&18.80\\
18060&21.42&20.58&19.58&18.80&18.40&18.48&18.12&17.58&17.42&17.17\\
18061&21.60&20.51&19.38&18.51&18.05&18.07&17.72&17.33&17.09&16.93\\
18062&22.04&21.36&20.56&19.90&19.65&19.59&19.34&18.71&18.40&18.06\\
25001&21.69&20.47&19.25&18.38&17.87&17.90&17.53&17.06&16.77&16.64\\
00002&22.09&20.55&19.25&18.31&17.82&17.88&17.53&17.15&16.81&16.68\\
00600&22.57&21.77&20.86&20.11&19.73&19.80&19.46&18.76&18.85&18.56\\
12004&21.87&21.19&20.50&19.86&19.50&19.44&19.26&$-9.00$&$-9.00$&$-9.00$\\
00007&20.79&20.25&19.62&19.05&18.73&18.76&18.24&18.33&18.14&17.98\\
12007&21.93&21.01&20.03&19.26&18.80&18.84&18.46&18.09&17.75&17.71\\
25100&21.08&20.39&19.58&18.93&18.50&18.49&18.22&$-9.00$&$-9.00$&$-9.00$\\
25010&23.10&22.12&21.08&20.23&19.70&19.87&19.47&18.77&18.30&18.04\\
00700&22.68&21.79&20.82&19.95&19.61&19.66&19.18&18.86&18.64&18.60\\
00014&22.77&21.73&20.57&19.59&19.26&19.23&18.83&18.09&17.73&17.57\\
25702&21.32&20.28&19.12&18.27&17.72&17.78&17.42&$-9.00$&$-9.00$&$-9.00$\\
00104&22.31&20.91&19.62&18.72&18.24&18.28&17.89&17.50&17.20&17.24\\
00020&21.38&20.63&19.84&19.14&18.72&18.83&18.51&17.89&17.66&17.65\\
00022&21.83&21.22&20.42&19.77&19.57&19.53&19.33&18.54&18.25&18.08\\
25015&22.35&21.39&20.37&19.59&19.11&19.03&18.72&$-9.00$&$-9.00$&$-9.00$\\
00026&22.65&21.97&20.92&19.93&19.30&19.35&18.94&18.30&17.93&17.82\\
00027&21.40&20.49&19.47&18.63&18.20&18.18&17.84&17.26&16.98&16.85\\
00028&22.83&21.33&20.03&19.07&18.62&18.61&18.21&17.69&17.42&17.39\\
12027&21.80&20.52&19.33&18.46&17.99&17.96&17.59&$-9.00$&$-9.00$&$-9.00$\\
00031&22.19&21.46&20.57&19.85&19.49&19.68&19.26&18.69&18.41&18.36\\
00034&21.68&20.27&18.99&18.07&17.56&17.62&17.23&16.68&16.38&16.32\\
00039&21.42&19.96&18.64&17.70&17.16&17.23&16.80&16.24&15.95&15.90\\
00201&21.96&21.30&20.52&19.87&19.61&19.64&19.34&18.74&18.55&18.57\\
12033&21.05&20.19&19.23&18.45&18.05&17.95&17.67&17.15&16.84&16.79\\
00047&20.76&20.20&19.55&18.92&18.67&18.67&18.40&17.80&17.47&17.38\\
00703&23.08&21.70&20.51&19.54&19.30&19.31&18.94&18.25&17.98&17.94\\
00051&22.04&21.34&20.57&19.94&19.61&19.74&19.38&18.84&18.78&18.80\\
00057&21.20&19.72&18.40&17.42&16.90&16.94&16.54&16.00&15.69&15.66\\
00105&21.25&19.78&18.42&17.47&16.92&16.99&16.54&16.00&15.70&15.65\\
00106&21.74&20.30&18.90&18.01&17.46&17.49&17.11&16.75&16.43&16.44\\
00060&21.29&20.44&19.50&18.74&18.31&18.34&17.99&17.48&17.23&17.13\\
00701&22.36&21.19&19.94&18.99&18.55&18.60&18.16&17.49&17.19&17.13\\
25701&21.86&20.50&19.25&18.35&17.84&17.83&17.53&17.11&16.86&16.70\\
12068&22.34&21.77&20.94&20.13&19.93&19.96&20.20&19.30&19.15&18.77\\
00077&22.81&21.87&20.93&20.15&20.09&19.97&19.64&18.93&18.55&18.29\\
06008&21.02&19.62&18.35&17.46&16.91&16.95&16.48&$-9.00$&$-9.00$&$-9.00$\\
12058&23.21&21.85&20.64&19.68&19.18&19.10&18.81&18.33&17.73&18.20\\
12100&23.30&22.18&21.06&20.20&19.78&19.72&19.56&18.94&18.40&18.19\\
00084&23.57&22.53&21.63&20.84&20.44&20.44&20.19&19.46&19.06&18.95\\
00085&21.83&21.16&20.39&19.78&19.58&19.57&19.30&18.83&18.63&18.63\\
12702&20.89&20.21&19.48&18.83&18.44&18.38&18.20&$-9.00$&$-9.00$&$-9.00$\\
00089&22.20&20.91&19.47&18.73&18.25&18.18&17.85&17.30&17.05&17.03\\
00704&21.74&20.68&19.52&18.60&18.13&18.19&17.79&17.21&16.92&16.94\\
12061&22.54&21.65&20.68&19.93&19.45&19.49&19.15&-9.00&-9.00&-9.00\\
00705&22.51&21.18&19.96&19.04&18.61&18.58&18.23&17.63&17.33&17.35\\
00098&22.30&21.25&20.16&19.29&18.84&18.77&18.43&17.81&17.52&17.43\\
13700&22.08&21.30&20.37&19.61&19.21&19.25&18.86&18.39&18.05&18.02\\
19011&22.60&21.51&20.38&19.46&18.97&18.88&18.58&17.83&17.42&17.35\\
13003&21.89&21.02&20.06&19.25&18.78&18.78&18.40&$-9.00$&$-9.00$&$-9.00$\\
13004&22.09&21.17&20.17&19.31&18.68&18.70&18.26&$-9.00$&$-9.00$&$-9.00$\\
13008&21.67&20.89&20.01&19.27&18.91&18.87&18.58&18.02&17.77&17.64\\
13015&21.89&20.45&19.15&18.19&17.69&17.69&17.28&17.03&16.75&16.58\\
13702&22.33&20.98&19.75&18.82&18.30&18.31&17.89&$-9.00$&$-9.00$&$-9.00$\\
19061&21.34&20.68&20.00&19.42&19.05&18.88&18.80&$-9.00$&$-9.00$&$-9.00$\\
\enddata
\end{deluxetable}

\clearpage

\begin{deluxetable}{lrrrrrlll}
\tablecolumns{9}
\tablewidth{0pt}
\tablecaption{A2125 Member Physical Parameters\label{pp}}
\tablenum{4}
\pagestyle{empty}
\tablehead{
\colhead{Name} & 
\colhead{Size} &
\colhead{$\log(L_{20cm}$)} &
\colhead{Proj Dist} &
\colhead{$M_R$} &
\colhead{Spec T.} &
\colhead{Age} &
\colhead{$A_V$} &
\colhead{z} \\
\colhead{} &
\colhead{kpc} &
\colhead{W/Hz} &
\colhead{Mpc} &
\colhead{mag} &
\colhead{} &
\colhead{Gyrs} &
\colhead{mag} &
\colhead{}} 
\startdata
17017&$<$8.4&22.40&3.7&-22.1&1&2.6&0.8&0.2475\\
23009&15.7&22.70&3.43&-21.6&2&5.5&1.2&0.2575\\
23016&95.8&25.00&2.93&-21.8&2&6.5&0.6&0.2449\\
18013&$<$5.0&22.41&2.38&-20.5&1&2.0&0.2&0.2449e\\
11700&$<$9.6&21.95&2.40&-20.7&2&6.5&0.0&0.2442e\\
18023&$<$4.6&22.51&2.12&-21.9&1&2.0&1.0&0.2448\\
24007&3.8&22.79&3.00&-21.2&2&5.5&0.4&0.2468e\\
24012&6.1&22.46&2.48&-21.3&3&9.5&0.6&0.2445e\\
18700&$<$6.9&21.82&1.74&-22.4&2&9.5&0.2&0.2458\\
24015&10.0&22.28&2.00&-21.4&2&4.5&0.2&0.2408e\\
18030&$<$3.4&22.15&1.76&-20.8&1&1.7&0.8&0.2482e\\
24016&13.8&22.66&2.07&-22.6&3&11.5&0.2&0.2458\\
18033&7.7&22.77&1.79&-21.8&5&11.5&0.4&0.2458e\\
18072&379.2&24.25&1.54&-22.4&1&7.5&0.2&0.2459\\
24027&17.6&23.21&1.83&-22.5&1&4.5&0.6&0.2425\\
24030&7.3&22.14&1.81&-20.8&2&4.5&0.0&0.2436e\\
18041&8.8&22.82&1.54&-22.9&1&11.5&0.0&0.2455\\
18701&32.6&22.40&1.32&-21.8&2&5.5&0.6&0.2477\\
18042&$<$2.7&22.74&1.50&-22.4&2&9.5&0.2&0.2458\\
24111&21.1&22.26&1.74&-20.8&2&1.7&0.4&0.2460e\\
18045&6.1&22.02&1.50&-21.1&1&0.1&0.8&0.2458e\\
18047&8.4&22.23&1.28&-21.1&4&5.5&0.0&0.2433e\\
18048&6.5&22.75&1.55&-21.9&5&11.5&0.2&0.2455e\\
24701&$<$7.3&21.83&1.60&-22.0&2&8.5&0.2&0.2559\\
18702&10.3&21.80&1.51&-21.6&3&6.5&0.0&0.2443e\\
24033&$<$7.3&21.91&1.60&-22.4&1&5.5&0.4&0.2565\\
18703&9.6&21.75&1.05&-22.0&2&8.5&0.2&0.2500\\
05012&6.9&22.86&3.06&-21.8&2&5.5&0.6&0.2573\\
18070&5.7&21.93&0.98&-20.0&3&3.5&0.2&0.2475e\\
24115&$<$5.0&21.98&2.04&-21.8&1&8.5&0.0&0.2421\\
18056&3.4&22.52&1.19&-21.3&3&7.5&0.6&0.2456e\\
18057&4.6&22.40&1.09&-22.2&2&7.5&0.4&0.2494e\\
18300&$<$7.7&21.69&1.15&-21.0&2&2.6&0.2&0.2458e\\
18060&10.0&22.52&1.14&-21.9&2&4.5&0.4&0.2474e\\
18061&11.9&22.22&0.95&-22.2&3&11.5&0.0&0.2430\\
18062&7.3&22.40&0.91&-20.7&3&1.0&1.6&0.2444e\\
25001&18.8&22.38&1.68&-22.3&1&2.0&0.6&0.2530e\\
00002&$<$3.4&22.42&0.89&-22.4&1&8.5&0.0&0.2548\\
00600&$<$5.7&21.82&1.07&-20.6&4&10.5&0.2&0.2423e\\
12004&6.5&22.06&1.78&-20.9&4&9.5&0.0&0.2381e\\
00007&10.3&22.40&0.92&-21.6&3&5.5&0.0&0.2462e\\
12007&11.9&22.10&1.05&-21.4&2&4.5&0.4&0.2429e\\
25100&18.4&22.55&2.35&-21.8&5&10.5&0.6&0.2521e\\
25010&$<$2.7&22.65&1.31&-20.5&2&2.3&1.6&0.2558\\
00700&8.0&21.83&0.59&-20.6&3&9.5&0.0&0.2469\\
00014&21.1&22.49&0.81&-20.9&2&4.5&1.0&0.2423\\
25702&42.1&22.85&2.19&-22.4&3&10.5&0.4&0.2561\\
00104&95.8&24.43&0.76&-22.0&1&7.5&0.0&0.2433\\
00020&17.2&22.39&0.57&-21.6&4&8.5&0.4&0.2419e\\
00022&5.7&22.50&0.84&-20.8&2&0.7&1.6&0.2588e\\
25015&$<$3.8&22.41&2.16&-21.1&2&4.5&0.6&0.2508e\\
00026&$<$2.7&22.55&1.01&-20.9&4&10.5&1.0&0.2484e\\
00027&3.1&22.72&0.50&-22.1&2&4.5&0.6&0.2460e\\
00028&$<$4.2&21.97&0.44&-21.5&1&7.5&0.2&0.2486\\
12027&11.9&22.71&2.18&-22.2&1&2.0&0.6&0.2463\\
00031&10.3&22.21&0.40&-20.7&2&2.3&1.0&0.2545e\\
00034&$<$3.8&21.85&0.68&-22.6&1&3.5&0.4&0.2395\\
00039&$<$3.1&22.54&0.30&-23.0&1&5.5&0.4&0.2457\\
00201&$<$20.7&22.09&0.57&-20.6&2&2.6&0.6&0.2454e\\
12033&16.1&22.32&0.78&-22.1&4&11.5&0.4&0.2406e\\
00047&53.6&24.59&0.14&-21.7&8&3.5&1.0&0.2528e\\
00703&24.1&22.42&0.17&-20.9&1&1.4&0.8&0.2494\\
00051&9.6&21.96&0.65&-20.6&4&9.5&0.0&0.2473e\\
00057&10.7&23.69&0.03&-23.3&1&5.5&0.4&0.2518\\
00105&15.3&23.06&0.03&-23.1&1&7.5&0.4&0.2470\\
00106&88.1&24.42&0.02&-22.7&1&9.5&0.0&0.2466\\
00060&11.1&22.27&0.34&-22.0&3&7.5&0.4&0.2422e\\
00701&$<$8.8&21.61&0.13&-21.6&3&11.5&0.4&0.2480\\
25701&$<$5.0&21.74&1.47&-22.4&1&4.5&0.2&0.2454\\
12068&$<$5.4&22.05&1.10&-20.5&2&2.3&0.6&0.2484e\\
00077&6.1&22.35&0.44&-20.2&1&0.2&1.8&0.2516e\\
06008&$<$4.6&22.77&3.06&-23.1&1&2.3&0.8&0.2369\\
12058&$<$3.4&22.40&1.43&-21.0&1&5.5&0.2&0.2514\\
12100&11.1&22.24&1.43&-20.5&2&4.5&0.8&0.2488e\\
00084&4.6&22.14&0.52&-19.8&2&2.6&1.2&0.2397e\\
00085&14.2&22.39&0.67&-20.9&2&2.6&0.4&0.2517\\
12702&$<$11.9&22.04&2.30&-21.9&5&11.5&0.4&0.2398e\\
00089&5.4&22.41&0.57&-21.8&2&8.5&0.2&0.2518\\
00704&31.0&22.70&0.64&-22.0&3&11.5&0.2&0.2547\\
12061&$<$3.4&22.40&2.06&-20.9&2&4.5&0.4&0.2492e\\
00705&35.6&22.51&0.78&-21.6&1&2.3&0.6&0.2444\\
00098&$<$3.4&22.10&0.99&-21.3&2&4.5&0.8&0.2473e\\
13700&$<$5.4&21.77&1.12&-20.9&4&10.5&0.4&0.2534\\
19011&17.6&22.07&0.94&-21.2&2&4.5&1.0&0.2412e\\
13003&10.7&22.48&2.36&-21.4&4&9.5&0.8&0.2468e\\
13004&16.5&22.89&2.31&-21.5&5&7.5&1.8&0.2458e\\
13008&12.3&22.18&1.29&-21.4&4&9.5&0.4&0.2501e\\
13015&5.0&22.59&1.59&-22.4&1&9.5&0.0&0.2473\\
13702&25.3&22.58&2.18&-21.8&1&2.0&0.8&0.2390\\
19061&7.7&22.34&2.29&-21.4&5&11.5&0.2&0.2394e\\
\enddata
\end{deluxetable}

\clearpage

\begin{deluxetable}{llllllll}
\tablecolumns{8}
\tablewidth{0pt}
\tablecaption{A2125 Radio Members with X-ray Detections   \label{XR}}
\tablenum{5}
\tablehead{
\colhead{Name} & 
\colhead{RA(2000.0)} &
\colhead{Dec(2000.0)} &
\colhead{X Name} &
\colhead{$X_{lum}$} &
\colhead{$R_{lum}$} &
\colhead{Abs mag} &
\colhead{z}} 
\startdata
24016&15 39 52.60&66 09 54.1&X014&41.9&22.66&-22.6&0.2458\\ 
24027&15 40 05.34&66 10 12.9&X022&41.7&23.21&-22.5&0.2425\\ 
18041&15 40 09.11&66 12 17.1&X025&41.7&22.82&-22.9&0.2455\\ 
24701&15 40 16.49&66 10 39.7&X033&41.7&21.83&-22.0&0.2559\\ 
00034&15 41 00.37&66 19 02.8&X063&41.6&21.85&-22.6&0.2395\\ 
00039&15 41 01.94&66 16 26.5&X065&42.1&22.54&-23.0&0.2457\\ 
00047&15 41 09.73&66 15 44.5&X068&41.7&24.59&-21.7&0.2528e\\ 
00704&15 41 43.39&66 14 19.1&X087&41.7&22.70&-22.0&0.2547\\ 
\enddata
\tablecomments{col 1: radio source name, col 2-3 R.A., Dec (2000.0),
col 4: X-ray source name \citep{w03}, col 5: log $0.5-8.0$ keV
X-ray luminosity, col 6: log 20cm radio
luminosity,
col 7: absolute R magnitude; col 8: redshift \citep{m03}}
\end{deluxetable}
\clearpage
\begin{deluxetable}{lllllllll}
\tablecolumns{9}
\tablewidth{0pt}
\tablecaption{A2125 Radio Members with Faint X-ray Detections.
   \label{FXR}}
\tablenum{6}
\tablehead{
\colhead{Name} & 
\colhead{RA(2000.0)} &
\colhead{Dec(2000.0)} &
\colhead{$X_{lum}$} &
\colhead{X band} &
\colhead{X class} &
\colhead{$R_{lum}$} &
\colhead{Abs mag} &
\colhead{z}} 
\startdata
18030&15 39 52.47&66 13 18.5&41.1&S&d&22.15&-20.8&0.2482e\\ 
18700&15 39 50.14&66 14 54.7&41.4&B&p&21.82&-22.4&0.2458\\  
18701&15 40 10.34&66 15 46.2&40.9&S&p&22.40&-21.8&0.2477\\  
18045&15 40 13.98&66 11 50.7&41.3&B&e&22.02&-21.1&0.2458e\\ 
18048&15 40 15.84&66 11 09.9&40.9&S&?&22.75&-21.9&0.2455e\\ 
18702&15 40 17.14&66 11 15.9&41.2&B&?&21.80&-21.6&0.2443e\\ 
24033&15 40 21.42&66 10 12.1&41.3&B&e&21.91&-22.4&0.2565\\  
00022&15 40 52.63&66 12 30.4&40.8&B&p&22.50&-20.8&0.2588e\\ 
00026&15 40 54.49&66 11 27.4&41.3&S&p&22.55&-20.9&0.2484e\\ 
00027&15 40 54.67&66 17 15.7&41.3&B&p&22.72&-22.1&0.2460e\\ 
00051&15 41 11.59&66 12 47.9&40.9&S&?&21.96&-20.6&0.2473e\\ 
00057&15 41 14.37&66 15 57.1&41.6&B&d&23.69&-23.3&0.2518\\  

\enddata
\tablecomments{col 1: radio source name, col 2-3 R.A., Dec (2000.0),
col 4: log $0.5-8.0$ keV
X-ray luminosity, col 5: log 20cm radio
luminosity, col 6: X-ray band in which detection was made:
S:soft(0.5--2.0 keV), B:both(0.5--8.0 keV), col 7:
detection class: d: diffuse feature, e: clearly extended,
p: point-like, ?: too faint to make a classification,
col 8: absolute R magnitude, col 9: redshift \citep{m03}}

\end{deluxetable}
\clearpage
\begin{deluxetable}{lllllll}
\tablecolumns{5}
\tablewidth{0pt}
\tablecaption{A2125 Members with X-ray Detections and no Radio Detection   \label{XNR}}
\tablenum{7}
\tablehead{
\colhead{Name} & 
\colhead{RA(2000.0)} &
\colhead{Dec(2000.0)} &
\colhead{$X_{lum}$} &
\colhead{z}} 
\startdata
X064&15 41 02.01&66 17 20.9&42.5&0.2554\\
X072&15 41 17.32&66 19 23.7&41.9&0.2463\\
\enddata
\tablecomments{col 1: X-ray source name\citep{w03} , 
col 2-3 R.A., Dec (2000.0),
col 4: log $0.5-8.0$ keV
col 5: redshift \citep{m03}}
\end{deluxetable}
\clearpage

\begin{deluxetable}{llllllll}
\tablecolumns{8}
\tablewidth{0pt}
\tablecaption{A2125 Star-formation Index\label{SFI}}
\tablenum{8}
\pagestyle{empty}
\tablehead{
\colhead{Name} & 
\colhead{$L_{20cm}$} &
\colhead{Align} &
\colhead{Mag} &
\colhead{Emis} &
\colhead{Age} &
\colhead{Dust} &
\colhead{Total}}
\startdata
17017&0.5&0.0&0.0&0.0&0.5&0.5&1.5\\
23009&0.0&1.0&0.0&0.0&0.0&1.0&2.0\\
23016&0.0&0.0&0.0&0.0&0.5&0.0&0.5\\
18013&0.5&0.0&0.5&1.0&0.5&0.0&2.5\\
11700&0.5&0.0&0.5&1.0&0.0&0.0&2.0\\
18023&0.5&0.0&0.0&0.0&0.5&1.0&2.0\\
24007&0.0&1.0&0.0&1.0&0.0&0.0&2.0\\
24012&0.5&1.0&0.0&1.0&0.0&0.5&3.0\\
18700&0.5&0.0&0.0&0.0&0.0&0.0&0.5\\
24015&0.5&1.0&0.0&1.0&0.5&0.0&3.0\\
18030&0.5&0.0&0.5&1.0&0.5&0.5&3.0\\
24016&0.5&0.0&0.0&0.0&0.0&0.0&0.5\\
18033&0.0&1.0&0.0&1.0&0.0&0.0&2.0\\
18072&0.0&0.0&0.0&0.0&0.0&0.0&0.0\\
24027&0.0&0.0&0.0&0.0&0.5&0.5&1.0\\
24030&0.5&1.0&0.5&1.0&0.5&0.0&3.5\\
18041&0.0&0.0&0.0&0.0&0.0&0.0&0.0\\
18701&0.5&1.0&0.0&0.0&0.0&0.5&2.0\\
18042&0.0&0.0&0.0&0.0&0.0&0.0&0.0\\
24111&0.5&1.0&0.5&1.0&0.5&0.0&3.5\\
18045&0.5&0.0&0.0&1.0&1.0&0.5&3.0\\
18047&0.5&1.0&0.0&1.0&0.0&0.0&2.5\\
18048&0.0&1.0&0.0&1.0&0.0&0.0&2.0\\
24701&0.5&0.0&0.0&0.0&0.0&0.0&0.5\\
18702&0.5&1.0&0.0&1.0&0.0&0.0&2.5\\
24033&0.5&0.0&0.0&0.0&0.0&0.0&0.5\\
18703&0.5&1.0&0.0&0.0&0.0&0.0&1.5\\
05012&0.0&1.0&0.0&0.0&0.0&0.5&1.5\\
18070&0.5&0.0&0.5&1.0&0.5&0.0&2.5\\
24115&0.5&0.0&0.0&0.0&0.0&0.0&0.5\\
18056&0.5&1.0&0.0&1.0&0.0&0.0&2.5\\
18057&0.5&1.0&0.0&1.0&0.0&0.0&2.5\\
18300&0.5&0.0&0.5&1.0&0.5&0.0&2.5\\
18060&0.5&1.0&0.0&1.0&0.5&0.0&3.0\\
18061&0.5&0.0&0.0&0.0&0.0&0.0&0.5\\
18062&0.5&0.0&0.5&1.0&1.0&1.0&4.0\\
25001&0.5&0.0&0.0&1.0&0.5&0.5&2.5\\
00002&0.5&0.0&0.0&0.0&0.0&0.0&0.5\\
00600&0.5&0.0&0.5&1.0&0.0&0.0&2.0\\
12004&0.5&1.0&0.5&1.0&0.0&0.0&3.0\\
00007&0.5&1.0&0.0&1.0&0.0&0.0&2.5\\
12007&0.5&1.0&0.0&1.0&0.5&0.0&3.0\\
25100&0.5&0.0&0.0&1.0&0.0&0.5&2.0\\
25010&0.5&0.0&0.5&0.0&0.5&1.0&2.5\\
00700&0.5&0.0&0.5&0.0&0.0&0.0&1.0\\
00014&0.5&1.0&0.5&0.0&0.5&1.0&3.5\\
25702&0.0&0.0&0.0&0.0&0.0&0.0&0.0\\
00104&0.0&0.0&0.0&0.0&0.0&0.0&0.0\\
00020&0.5&1.0&0.0&1.0&0.0&0.0&2.5\\
00022&0.5&0.0&0.5&1.0&1.0&1.0&4.0\\
25015&0.5&0.0&0.0&1.0&0.5&0.5&2.5\\
00026&0.5&0.0&0.5&1.0&0.0&1.0&3.0\\
00027&0.0&0.0&0.0&1.0&0.5&0.5&2.0\\
00028&0.5&0.0&0.0&0.0&0.0&0.0&0.5\\
12027&0.0&1.0&0.0&0.0&0.5&0.5&2.0\\
00031&0.5&1.0&0.5&1.0&0.5&1.0&4.5\\
00034&0.0&0.0&0.0&0.0&0.5&0.0&1.0\\
00039&0.5&0.0&0.0&0.0&0.0&0.0&0.5\\
00201&0.5&0.0&0.5&1.0&0.5&0.5&3.0\\
12033&0.5&1.0&0.0&1.0&0.0&0.0&2.5\\
00047&0.0&1.0&0.0&1.0&0.5&1.0&3.5\\
00703&0.5&0.0&0.5&0.0&0.5&0.5&2.0\\
00051&0.5&0.0&0.5&1.0&0.0&0.0&2.0\\
00057&0.0&0.0&0.0&0.0&0.0&0.0&0.0\\
00105&0.0&0.0&0.0&0.0&0.0&0.0&0.0\\
00106&0.0&0.0&0.0&0.0&0.0&0.0&0.0\\
00060&0.5&1.0&0.0&1.0&0.0&0.0&2.5\\
00701&0.5&0.0&0.0&0.0&0.0&0.0&0.5\\
25701&0.5&0.0&0.0&0.0&0.5&0.0&1.0\\
12068&0.5&0.0&0.5&1.0&0.5&0.5&3.0\\
00077&0.5&1.0&0.5&1.0&1.0&0.5&5.0\\
06008&0.0&0.0&0.0&0.0&0.5&0.5&1.0\\
12058&0.5&0.0&0.0&0.0&0.0&0.0&0.5\\
12100&0.5&0.0&0.5&1.0&0.5&0.5&3.0\\
00084&0.5&1.0&0.5&1.0&0.5&1.0&4.5\\
00085&0.5&0.0&0.5&0.0&0.5&0.0&1.5\\
12702&0.5&0.0&0.0&1.0&0.0&0.0&1.5\\
00089&0.5&0.0&0.0&0.0&0.0&0.0&0.5\\
00704&0.0&0.0&0.0&0.0&0.0&0.0&0.0\\
12061&0.5&0.0&0.5&1.0&0.5&0.0&2.5\\
00705&0.5&0.0&0.0&0.0&0.5&0.5&1.5\\
00098&0.5&0.0&0.0&1.0&0.5&0.5&2.5\\
13700&0.5&0.0&0.5&0.0&0.0&0.0&1.0\\
19011&0.5&1.0&0.0&1.0&0.5&1.0&4.0\\
13003&0.5&1.0&0.0&1.0&0.0&0.5&3.0\\
13004&0.0&1.0&0.0&1.0&0.0&1.0&3.0\\
13008&0.5&1.0&0.0&1.0&0.0&0.0&2.5\\
13015&0.5&0.0&0.0&0.0&0.0&0.0&0.5\\
13702&0.5&1.0&0.0&0.0&0.5&0.5&2.5\\
19061&0.5&1.0&0.0&0.0&0.0&0.0&1.5\\
\enddata
\end{deluxetable}
\end{document}